\documentclass[11pt]{article}

\usepackage{setup}
\usepackage[para]{footmisc}
\usepackage{parskip}
\usepackage[backend=biber, style=numeric, maxnames=3, minnames=3, url=false]{biblatex}
\usepackage{microtype}
\usepackage{bbm}

\renewcommand{\mathbb}{\mathbf}

\newcommand{\rmv}{\!\!\!}
\usepackage{siunitx}
\usepackage{subcaption}

\newcommand\blfootnote[1]{%
  \begingroup
  \renewcommand\thefootnote{}\footnote{#1}%
  \addtocounter{footnote}{-1}%
  \endgroup
}

\title{Progress Towards Decoding Visual Imagery via fNIRS}

\author{
\blfootnote{\label{corresponding}\:\:\,\textsuperscript{$*$}Corresponding author. Email: raphael.hotter@gmail.com \\}
Michel Adami\v{c}\footnote{\label{mcgill-physics}\rmv McGill University, Department of Physics},
Wellington Avelino\textsuperscript{\ref{mcgill-physics}},
Anna Brandenberger\footnote{\label{mit-math}\rmv Massachusetts Institute of Technology, Department of Mathematics},
Bryan Chiang\footnote{\label{stanford}\rmv Stanford University, Department of Computer Science}, \\
Hunter Davis,
Stephen Fay\textsuperscript{\ref{mcgill-physics}},
Andrew Gregory,
Aayush Gupta,
Raphael Hotter\textsuperscript{\hyperref[corresponding]{$\ast$}},\\
Grace Jiang,
Fiona Leng,
Stephen Polcyn,
Thomas Ribeiro\footnote{\label{mcgill-engineering}\rmv McGill University, Department of Biomedical Engineering},
Paul S. Scotti\footnote{\rmv Princeton Neuroscience Institute}\footnote{\rmv Medical AI Research Center (MedARC)}\footnote{\rmv Stability AI},\\
Michelle Wang\footnote{\rmv McGill University, Department of Quantitative Life Sciences},
Marley Xiong,
Jonathan Xu\footnote{\rmv University of Waterloo, Department of Computer Science}
}

\begin{document}

\maketitle 

\abstract{We demonstrate the possibility of reconstructing images from fNIRS brain activity and start building a prototype to match the required specs. By training an image reconstruction model on downsampled fMRI data, we discovered that cm-scale spatial resolution is sufficient for image generation. We obtained 71\% retrieval accuracy with 1-cm resolution, compared to 93\% on the full-resolution fMRI, and 20\% with 2-cm resolution. With simulations and high-density tomography, we found that time-domain fNIRS can achieve 1-cm resolution, compared to 2-cm resolution for continuous-wave fNIRS. Lastly, we share designs for a prototype time-domain fNIRS device, consisting of a laser driver, a single photon detector, and a time-to-digital converter system.}{}

\section{Introduction} \label{sec:intro}
\begin{figure}
    \centering
    \includegraphics[scale=0.5]{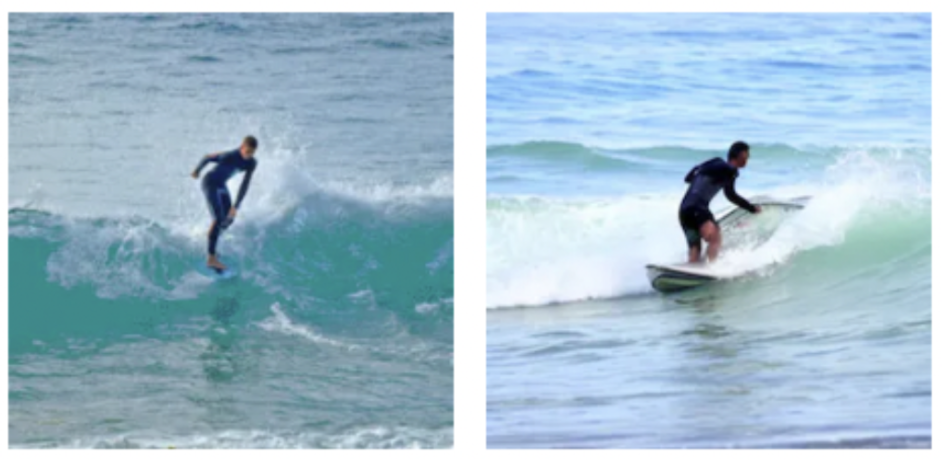}
    \caption{Left: image shown to participant. Right: image reconstructed from fMRI via MindEye.}
    \label{fig:mindeye}
\end{figure}

MindEye \cite{scotti2023reconstructing, scotti2024mindeye2} recently showed astonishing reconstructions of perceived images from brain activity. Functional magnetic resonance imaging (fMRI) data was collected while images were shown to participants, and then reconstructed with large multimodal models from brain activity, as shown in Figure~\ref{fig:mindeye}.

It seemed to us that new land had been breached. Could we build a device that travels with us, reads our visual representations, and allows us to conjure new images with our imaginations? Such a device would be a new interface for communication and would enable us to peer into those curious depths of the mind that are so far absent in our daily lives.

An fMRI machine is the size of a small vehicle, since supercooling infrastructure is required to generate magnetic fields strong enough to perturb tissue \cite{Ogawa1990}. The spatial resolution of fMRI is typically millimetres \cite{Logothetis2001}. On the other hand, electroencephalography (EEG) is small and portable, requiring as much hardware as a voltmeter. However, the resolution of EEG is limited to 2-3 cm \cite{Nunez2006}.\footnote{That said, in the last decade, work has shown that you might be able to get higher EEG resolution if you drastically increased the density of electrodes and reduced the noise floor. \cite{Grover2017, Robinson2017}}

Thankfully, there’s a modality that is portable and whose spatial resolution has yet to be pushed to its physical limits. You can measure a similar underlying signal as fMRI optically, rather than magnetically. Human tissue and water are largely transparent to infrared light, while blood is more absorbent. When you shine light through skin, you can track changes in absorbance. Since the absorption of blood depends on the local amount of oxygen, absorption changes can tell you about changes in blood oxygenation. This technique using light to measure blood oxygenation is called functional near-infrared spectroscopy, or fNIRS \cite{Ferrari2012}.

An fNIRS device is a set of light sources and detectors. There exist several forms of fNIRS. Naively, you could emit constant-intensity light and measure the intensity of light received at the scalp after it has traversed the cortex. This is considered continuous-wave, or CW-fNIRS \cite{Scholkmann2014}. Somewhat more exotically, you could measure the time-of-flight of photons. Time-domain or TD-fNIRS requires sub-nanosecond electronics for capturing photon arrival times \cite{Torricelli2014}. A TD device yields richer information about the paths that photons take.

Today, commercial CW devices exist and are on the  market (Lumo \cite{lumo}, NirX, TechEN, etc.). Time-domain hardware is more difficult to build, less commercially available, and more expensive; for example, a picosecond pulsed laser costs tens of thousands of dollars off the shelf.

Notably, with a portable and lower-cost device, we could collect orders of magnitude more training data than we can with an MRI machine. More data could greatly improve image generation and justify a potentially lower spatial resolution.

To ascertain whether fNIRS could viably be used in an image reconstruction device, we asked: what spatial resolution is required for image reconstruction? What spatial resolution is attainable with fNIRS? We then set out to build a TD-fNIRS device from scratch.

In this paper, we demonstrate that image reconstruction from brain activity appears feasible with 1 cm spatial resolution, using downsampled fMRI data (Section \ref{sec:resolution}). We share novel tomography simulations that show TD-fNIRS can achieve this resolution, and we’re the first, to our knowledge, to directly compare the spatial resolution achievable in CW and TD simulations (Section \ref{sec:tomography}). In Section \ref{sec:hardware}, we share our progress on a low-cost prototype for TD-fNIRS. Finally, in the Supplementary Information, we share our learnings from collecting fNIRS data from an existing CW headset, as well as more details on the hardware.

\section{Resolution needed for image reconstruction}\label{sec:resolution}
We took an fMRI dataset of people looking at images and downsampled it to various spatial resolutions. We then trained and evaluated MindEye on the downsampled dataset. We found that, despite roughly 125x fewer voxels, models trained on 1 cm resolution fMRI data can still achieve high retrieval accuracy.

\subsection{Dataset and model}
We used the Natural Scenes Dataset \cite{allen2022massive}, a public fMRI dataset containing brain responses of participants viewing natural scenes from MS-COCO \cite{lin2015microsoft}. The dataset used a 7T MRI machine to collect high-resolution (1.8-mm isotropic) whole-brain volumes. Eight participants were scanned for 30-40 hours each, comprising 22,000 to 30,000 trials of fMRI responses each, where fMRI responses correspond to session-wise z-scored single-trial betas output from GLMSingle \cite{10.7554/eLife.77599}.

Given that MindEye was trained and evaluated independently for each subject, we specifically trained our downsampled MindEye model on data from subject 1, the participant with the highest signal-to-noise ratio, and compared our results to the original MindEye model’s performance on subject 1.

We blurred and downsampled the original preprocessed fMRI volumes from 0.18 cm to 0.90 cm, 1.44 cm, and 1.98 cm resolution (which we refer to as 1 cm, 1.5 cm, and 2 cm). The original MindEye model only used voxels from the occipital region that were shown receptive to visual information, defined by the original NSD authors as the nsdgeneral brain region of interest. We likewise kept voxels that had some overlap with this mask.

Our downsampled MindEye model was then trained in the same manner as MindEye1, using their open-sourced code from GitHub \cite{fmri_reconstruction_nsd_2023}. MindEye1 is based on Stable Diffusion \cite{rombach2022highresolution} and consists of two submodules that are trained end-to-end and specialized for the tasks of retrieval (using contrastive learning with CLIP \cite{radford2021learning}) and reconstruction (using a diffusion prior trained from scratch).

\subsection{Results}
To evaluate retrieval, we compute the cosine similarity of the fMRI embedding with the CLIP embedding of the ground truth image as well as 299 random images from the test set. We repeated the procedure 30 times to account for variability in the images selected.

We found that 1-cm resolution achieves a 300-way top-1 retrieval accuracy of 71\%, compared to 93\% resulting from the original data (random chance = 0.3\%). Accuracy dropped off substantially below 1-cm resolution. Note that our performance on Subject 1 was lower than what MindEye reported (97\%), despite not modifying their architecture (we suspect this was a package dependency problem).

\begin{figure}
    \centering
     \includegraphics[width=\textwidth]{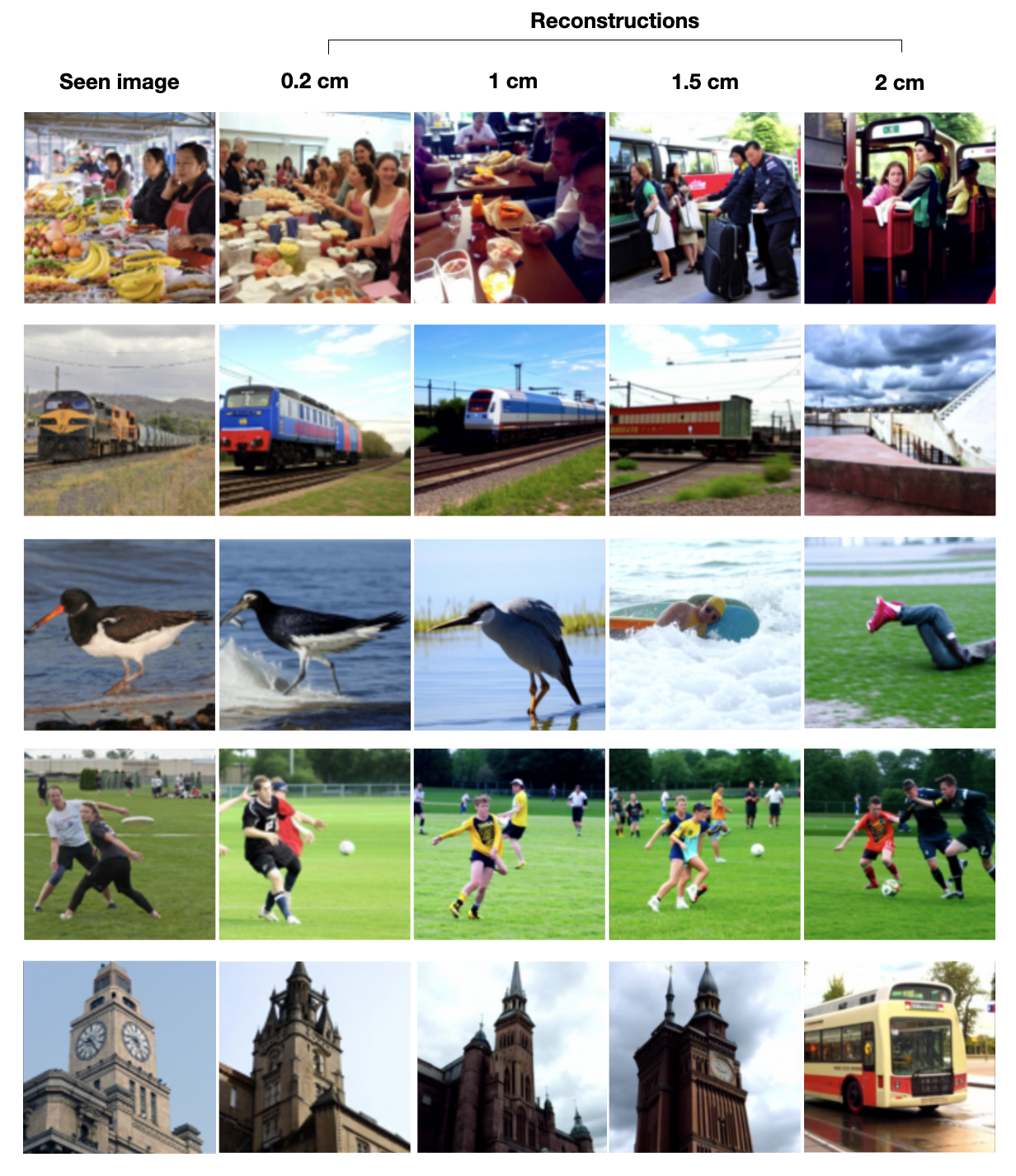}
    \caption{Example reconstructions from downsampled MindEye at various resolutions.}
    \label{fig:recons}
\end{figure}

We also used the downsampled models to generate image reconstructions. Qualitatively, we found that the 1-cm model could generate images comparable to the 1.8-mm model (Figure \ref{fig:recons}). Decreasing the resolution below 1 cm produced images of lower quality.

Therefore, 1-cm resolution seems necessary and potentially sufficient for reconstructing images from the brain.

\begin{table}[ht]
\centering
\caption{Retrieval performance of MindEye on downsampled fMRI. We trained and evaluated MindEye on downsampled versions of the NSD dataset.}
\begin{tabular}{cc}
\toprule
\textbf{Resolution} & \textbf{300-way Top 1 Retrieval Accuracy} \\ \midrule
0.18 cm             & 93\%                                      \\
1 cm                & 71\%\\ 
1.5 cm              & 33\%\\ 
2 cm                & 20\%                                      \\ \bottomrule
\end{tabular}
\label{tab:mind_eye_performance}
\end{table}

\subsection{Limitations}
Our investigations should only be treated as a first-order estimate of the resolution necessary to decode images from the brain. fNIRS is not equivalent to downsampled fMRI: though they are both based on the haemodynamic response, fMRI measures the $\text{T2}^*$ relaxation time of protons, whereas fNIRS measures the absorption of light. Importantly, the two modalities have different signal-to-noise ratios, potentially different temporal resolution (fNIRS could have higher), and different depth penetration (fNIRS is more limited). Furthermore, we did not optimize the hyperparameters of MindEye for low resolution data, and future models could improve results further. MindEye also flattens voxels to a 1-dimensional vector, which removes important spatial information that would be preserved using flattened 2D cortical surface maps \cite{chen2023structure} or native 3D and 4D voxel space \cite{kim2024swift}.

\section{Tomographic spatial resolution simulation}\label{sec:tomography}
After we got an estimate of the spatial resolution requirements, we wanted to know whether we could achieve this resolution with fNIRS. Surprisingly, this question has not been answered yet: White and Culver \cite{White2010} estimated the resolution of continuous-wave fNIRS, but no study to our knowledge has done so for time-domain fNIRS with realistic sensor geometries. Gao et al. \cite{Gao:02} compared TD and CW resolution in a fan-beam (transmission) geometry, but their results are not quantitative, and they did not use the reflection geometry that is used in fNIRS.

We ran simulations to estimate the resolution of time-domain and continuous-wave fNIRS. The relationship between sensor density and resolution is not straightforward; light scatters through the brain, and must be mapped back to an image of the object which created it in a process called tomography.

\subsection{Diffuse optical tomography}
To run simulations, we needed a model for how photons propagate in tissue. We use the radiative transfer equation (RTE) \cite{bigio2016quantitative}, which accounts for both scattering and absorption phenomena. The RTE can be written as
\begin{equation}
\frac{1}{v} \frac{\partial I(\mathbf{r}, \mathbf{\hat{s}}, t)}{\partial t} + 
\hat{\mathbf{s}} \cdot \nabla I(\mathbf{r}, \mathbf{\hat{s}}, t) = 
- [\mu_a(\mathbf{r}) + \mu_s(\mathbf{r})] I(\mathbf{r}, \mathbf{\hat{s}}, t) +
\mu_s(\mathbf{r}) \oint p(\mathbf{\hat{s}}, \mathbf{\hat{s}}') I(\mathbf{r}, \mathbf{\hat{s}}',t) d\Omega'
\end{equation}

where $I(\mathbf{r}, \mathbf{\hat{s}}, t)$ is the intensity of light at a position $\mathbf{r}$, in the direction $\mathbf{\hat{s}}$ and at time $t$, $v$ is the speed of light in tissue, $\mu_a$ and $\mu_s$ are the absorption and scattering coefficients, and $p(\mathbf{\hat{s}},\mathbf{\hat{s}'})$ is the scattering phase function. Each term in the equation has units of radiance, i.e. dimensions of  \si{\watt/\meter\squared \steradian}, so the RTE can be viewed as a form of conservation of energy.

Given a distribution of scattering and absorption parameters $\mu_a(\mathbf{r})$, $\mu_s(\mathbf{r})$, we can solve the RTE to find $I(\mathbf{r}, \mathbf{\hat{s}}, t)$, i.e. the paths that light takes in that medium. This is referred to as the forward problem.

The inverse problem is to recover the distribution of scattering and absorption parameters from samples of $I(\mathbf{r}, \mathbf{\hat{s}}, t)$ at the light detector locations. We frame the inverse problem as an optimization problem: 
\begin{equation}
\min_{\mu_a(\mathbf{r}), \mu_s(\mathbf{r})} \sum_{i,j} \left\| y_{ij} - \mathcal{F}_{ij}\Big[\mu_a(\mathbf{r}), \mu_s(\mathbf{r})\Big] \right\|_2^2
\end{equation}
where $y_{ij}$ and $\mathcal{F}_{ij}$ are the measured logarithms of the light intensity and simulated light intensity respectively located at source position $i$ and detector position $j$. The norm $\|.\|_2$ is the $l_2$ norm and sums over time. We can include a prior over maps of brain activity by adding a regularization term.

We solve the optimization problem iteratively: we make a guess for some $\mu_a$ and $\mu_s$, run a forward RTE solver to find the corresponding intensity at the detectors, compare our simulated intensity with our measurements, then update the guess so that the simulated intensity becomes closer to measured intensity.

\subsection{Simulation details}
We ran simulations on a simple medium: a two-dimensional optically-homogeneous disk with two localized inclusions (Figure \ref{fig:spatial_res}). We sought to reconstruct those inclusions by solving the inverse problem, varying (a) the distance of separation between the localized inclusions to see if our inverse solution could distinguish between them, and (b) the depth at which the inclusions were placed within the homogeneous medium. We ran these experiments for both time-domain and continuous-wave fNIRS.

We picked a radius of 7 cm for the disk, which is roughly the radius of a human head. We used realistic values for the absorption coefficient, scattering coefficient, and refractive index parameters in biological tissue: \(\mu_a = 0.02 \, \text{mm}^{-1}\), \(\mu_s = 0.67 \, \text{mm}^{-1}\), \(n = 1.4\) \cite{spinelli2017vivo}. We set the inclusion to have 2x greater absorption and scattering coefficient (we did not observe a difference in our results at 1.1x or 1.01x contrast). We used 10 sources and 10 detectors (higher density did not improve results). For TD, we used 7 time bins of 640 ps each (we did not see improvements above this time resolution). We defined the depth to be the distance to the top of the inclusion.

We solve the forward problem using the Toast++ software package \cite{schweiger2014toast}, which makes a diffusion approximation to the radiative transfer equation \cite{bigio2016quantitative,boas_et_al_2006}. We discretize the medium on a triangle mesh, with finer resolution near the boundary. For the inversion, we use a Gauss-Newton optimizer (20 iterations) with Conjugate Gradients (max of 100 inner iterations) and total variation regularization (\(\tau = 0.01\), \(\beta = 0.01\)). See \texttt{reconTD\_GN.m} (\href{https://github.com/neural-imagery/tomography-experiments/blob/257abb7212ef755786f83564d6865983b1cb9d06/recon_2obj/reconTD_GN.m}{source}) for the full details.

\subsection{Results and limitations}
\begin{figure}[p]
    \centering
    \includegraphics[width=\textwidth]{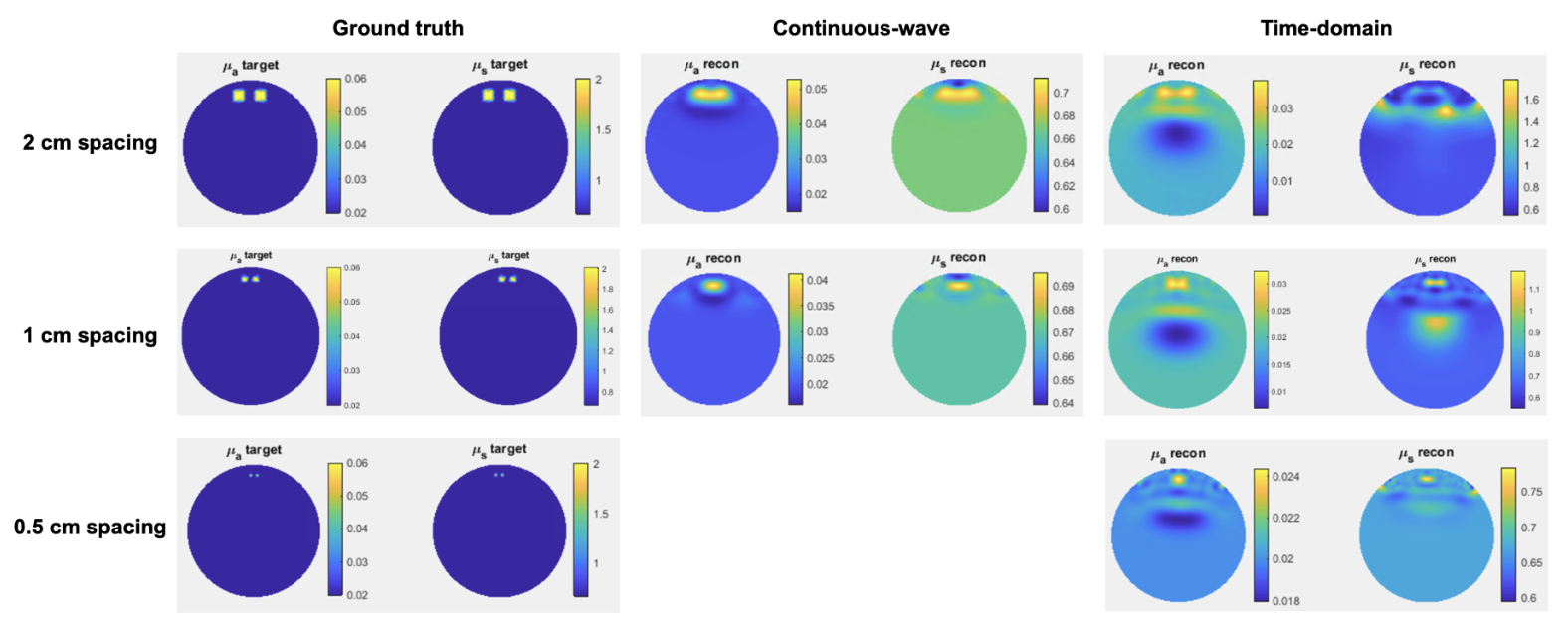}
    \caption{Spatial resolution of time-domain and continuous-wave fNIRS. The inclusions are placed 0.5 cm, 1 cm, and 2 cm apart. The width of each inclusion is set to be half of the separation.}
    \label{fig:spatial_res}
\end{figure}

\begin{figure}[p]
    \centering
    \includegraphics[width=\textwidth]{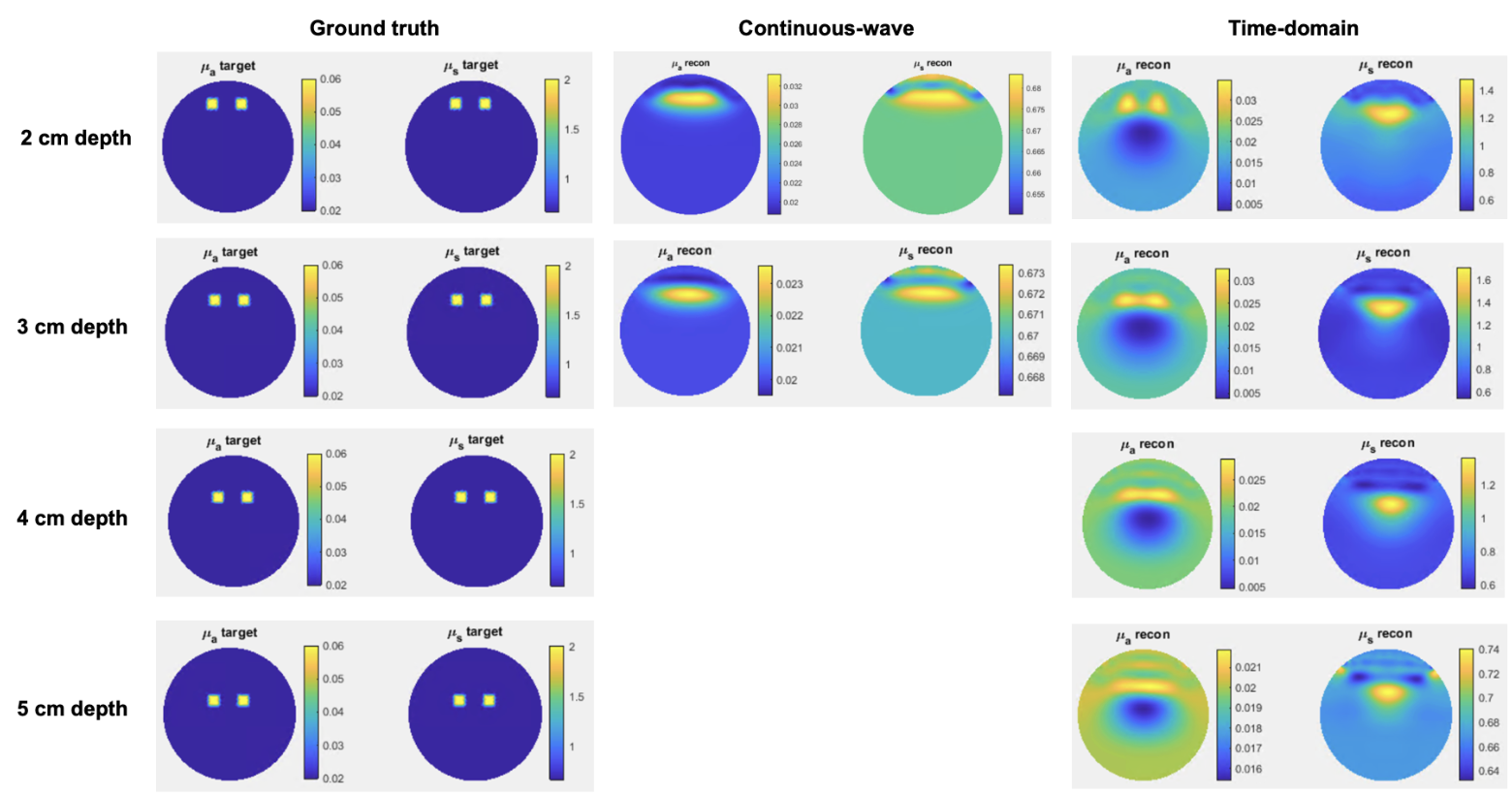}
    \caption{Depth penetration of time-domain and continuous-wave fNIRS. The centers of the inclusions are placed at a depth of 2 cm, 3 cm, 4 cm, and 5 cm. The inclusions are placed 3 cm apart.}
    \label{fig:depth}
\end{figure}

We find that, in a 2D setting, the spatial resolution of TD-fNIRS is significantly higher than CW. The minimum spacing for discernable reconstructions in \(\mu_a\) is 1 cm for TD, but only 2 cm for CW, at a depth of 1 cm (Figure \ref{fig:spatial_res}). Furthermore, TD can reconstruct inclusions at a depth of up to 4 cm for 3 cm spacing, but CW can only reach a depth of 2 cm for that spacing (Figure \ref{fig:depth}).

To our knowledge, these results are the first to show quantitatively that TD has better spatial resolution and depth penetration than CW. Together with the results from Section \ref{sec:resolution}, it seems a TD-fNIRS system might be able to reconstruct images from the brain.

There are many limitations in our simulations. We do not model the skull, hair, or any inhomogeneities in the optical parameters of the brain. We don’t include any noise or motion artifacts. Our simulation is 2D and not 3D. Other optimizers might get better reconstructions, especially if they include better priors over brain-like solutions, for example, via deep learning.

\section{Time-domain hardware prototype}\label{sec:hardware}

Our goal was to build a low-cost TD-fNIRS prototype. Though we did not complete a full prototype, in this section, we present an overview of our design, which includes a gain-switching laser, single photon detectors, and time-to-digital converter.

\subsection{System overview and requirements}
\begin{figure}[hbtp]
    \begin{center}
        \begin{subfigure}{0.55\textwidth}
            \includegraphics[width=\textwidth]{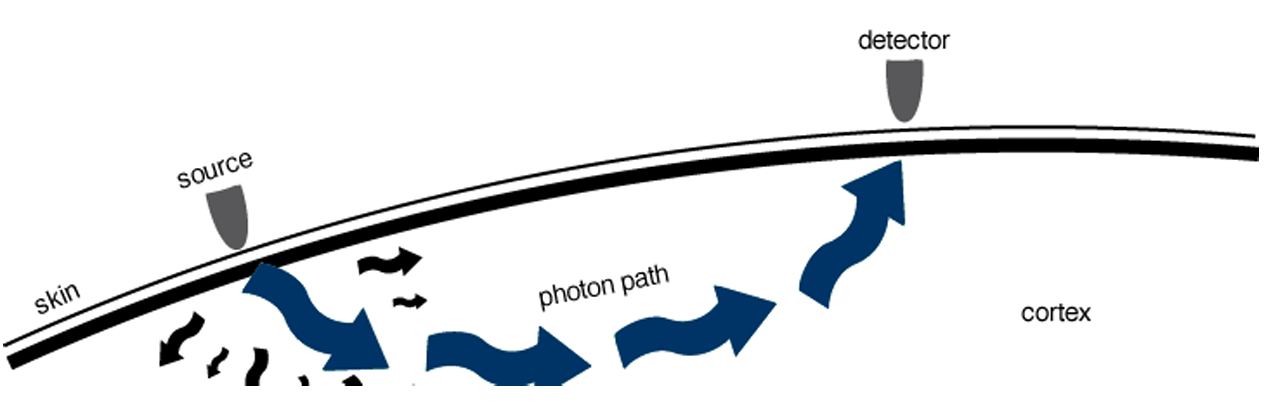}
        \end{subfigure}
        \begin{subfigure}{0.40\textwidth}
            \includegraphics[width=\textwidth]{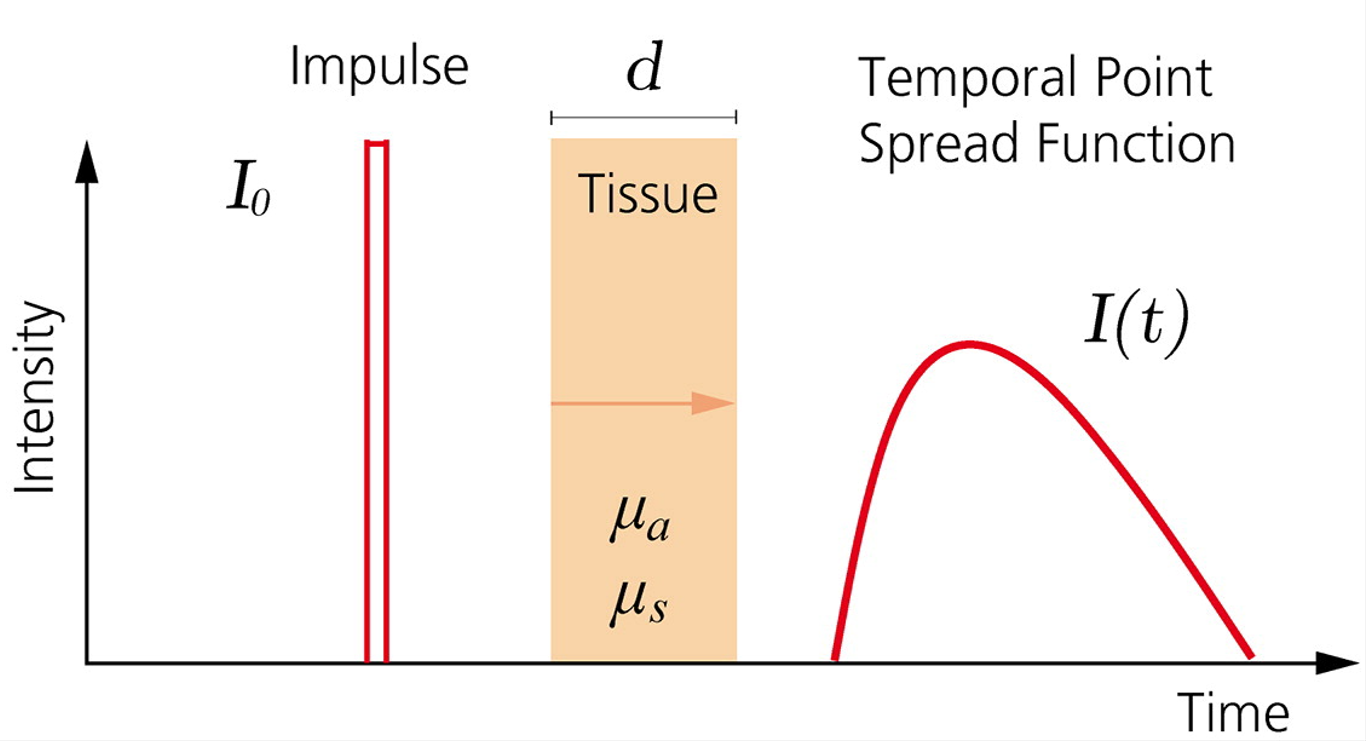}
        \end{subfigure}
    \end{center}
    \caption{Left: Photons follow a banana-shaped path through the tissue and are picked up by the detector a couple centimeters away from the source (image from \cite{Strait_2014}). Right: TD fNIRS emits a very short pulse of light in the tissue and measures the arrival times of photons that emerge (image from \cite{Scholkmann2014}).}
    \label{fig:TD_fNIRS}
\end{figure}

In TD-fNIRS, a short pulse (\textasciitilde 100 ps duration) of monochromatic near-infrared light is emitted into the brain and picked up by detectors placed on the scalp nearby (Figure \ref{fig:TD_fNIRS}). Since photons scatter and get absorbed while travelling through tissue, the measured light intensity will have a delayed, attenuated and broadened temporal distribution. The shape of this distribution, which is often called the temporal point spread function (TPSF) in literature \cite{Scholkmann2014, Torricelli2014}, contains information about the different paths light took while travelling through the brain. We can reconstruct these paths using tomography, and if we measure the TPSF at multiple wavelengths, the blood oxygenation inside the tissue can be determined.

In biological tissue, it takes near-infrared light an average of 1 nanosecond to travel from a source to a detector 3 cm away \cite{bigio2016quantitative}. This duration is about 8 times longer than the time it would take for light to travel 3 cm through pure water\footnote{This quantity is called the differential path length factor (DPF)} — the scattering makes the average pathlength longer. The width of the TPSF is also on the order of 1 nanosecond \cite{bigio2016quantitative}, so if we want to accurately sample the TPSF, the TD system's timing resolution should be on the order of 100 ps. This calls for very fast pulsed light sources, detectors and data acquisition systems.

How do we approach this from an engineering standpoint? One approach is to sample the detector output with a very high speed analog-to-digital converter (ADC). However, digitizing the signal every 100 ps, i.e., at 10 GHz speeds, is very expensive. For example, a Texas Instruments 10.4 Gsps analog-to-digital converter costs \$1,822  per detector \cite{TI_ADC12DJ5200SE}. Instead of sampling at 10 Gsps, we use a time-correlated single photon counting scheme. In this scheme, we tune the power of our laser so that each detector detects at most 1 photon per pulse, and we measure its arrival time using a time-to-digital converter (TDC). We then use many pulses to build a distribution of time of flights (DTOF) histogram, which corresponds to the temporal point spread function. Like this, we can achieve very high timing accuracy at the expense of a lower TPSF sampling rate.

How many photons per second does our system need to detect? The Kernel Flow system detects on the order of $10^7$ counts per histogram \cite{KernelFlow}. We take $10^7$ counts per histogram as a target for our design. Since we're measuring a haemodynamic process, the signal should only change significantly over a period of seconds, so we aim to sample a DTOF histogram roughly once per second. Therefore, we target a photon sampling rate of around 10 Msps per detector.

We aim for the system to have a high density of optodes -- White and Culver \cite{White2010} show that a high-density (1.3 cm minimum spacing between optodes) CW system achieves a significantly higher resolution than a low-density (3 cm spacing) system. The visual cortex is roughly 10 cm x 10 cm, so we estimate that our system would need roughly 100 optodes to cover the full visual cortex at high density.

In this section, we present our take to building low-cost subsystems which may fulfill these requirements, and the challenges that need to be addressed to have a full working demo.

\subsection{Pulsed light source}
To measure the time of flight of photons with such high timing precision, very narrow pulses of light need to be produced (below 100 ps). Semiconductor laser diodes are suitable for this, but driving them with such short electrical pulses is challenging due to the high frequencies involved (10 GHz). Our proposed solution lies in a method called gain switching, a technique used in lasers to generate short optical pulses, typically on the order of picoseconds, without needing sub-nanosecond injection currents \cite{Lau_1988}. The idea is to bias the laser diode with a DC current right below the lasing threshold, and then apply an AC square wave on top of it, exceeding the threshold every half-cycle. As this happens, stimulated emission quickly depopulates the excited states in the material, producing a very short burst of light, much shorter than the applied electrical pulse.

We decided to develop a pulsed laser driver circuit that would implement this with an off-the-shelf laser diode. Even though the driver performed as expected electrically, we unfortunately failed to observe any gain switching with our selected diode, although work is still underway to attempt to achieve it. We present the design of our laser driver circuit and some tests with a ThorLabs HL6738MG 690 nm laser diode in the Supplementary material in Section \ref{sub:laser_driver}.

\subsection{Single photon detectors}
Detecting single photons with high timing accuracy is not trivial, and we decided to investigate two different solutions --- a fast, actively quenched avalanche photodiode (APD), and a silicon photomultiplier (SiPM) detector. APDs and SiPMs are modern solid state devices, capable of fast single photon detection. They are in fact closely related --- SiPMs are just large arrays of single photon APDs (SPADs), allowing them to detect multiple photons at once. We discuss both approaches below.

\subsubsection{Avalanche photodiode (APD)}
When used in Geiger mode (i.e. with a bias above the breakdown voltage), APDs can detect single photons by producing an electrical pulse of sufficient amplitude to directly trigger a logic event. The device must then be reset by quenching the avalanche, which can be done either passively or actively \cite{Photodetectors, Stipcevic_2010}.

Passive quenching involves using a simple resistor in series with the APD. When a photon is absorbed and an avalanche is triggered, the current through the resistor increases, causing a voltage drop across the resistor that reduces the bias across the APD below the breakdown voltage. This quenches the avalanche. The bias then slowly recovers to the operating level, and the detector is ready to be triggered again. This mode is called passive quenching passive reset (PQPR). An active circuit element can alternatively be used for reset, forming a passive quench active reset (PQAR) mode. Passive quenching is simpler to implement in terms of circuitry, but suffers from longer detector dead times, which limits the maximum count rate.

Active quenching, on the other hand, involves a more complex circuit to quickly reduce the bias across the APD when an avalanche is detected. The bias is then rapidly restored to the operating level, reducing the dead time and allowing the detector to respond to incoming photons more quickly. This mode is called active quench active reset (AQAR). Active quenching can lead to shorter dead times compared to passive quenching, but is less straightforward to implement.

Since our application demands very high count rates for the reconstruction of DTOF histograms, we decided to build an active quenching circuit for our APD. It is based on the design proposed by M. Stipčević \cite{Stipcevic_2009} and we present its implementation and preliminary test results with a Hamamatsu S12023-02 Si APD in the Supplementary material in Section \ref{sub:apd_circuit}. 

The APD combined with the active quenching circuit had a dead time of about 40 ns, allowing for count rates up to around 20 Msps, which meets our requirements for the desired photon count rates. It came as a pleasant surprise that the chosen APD worked in Geiger mode, despite not being sold as a single photon APD (SPAD). Regular APDs are usually not optimized to operate in Geiger mode, but thanks to the active quenching circuit, we were able to demonstrate that a low cost APD can function as a fast single photon detector.

\subsubsection{Silicon photomultiplier (SiPM) detector}
SiPMs are modern solid state alternatives to photomultiplier tubes (PMTs) and are widely used for single photon detection. They consist of thousands of tiny SPADs (single photon pixels) with built-in quenching resistors. They are ideal for photon counting applications, which is technically not needed for our application, since we are only interested in single photons. Nevertheless, they are a popular choice for single photon detection as well.

We tested a SiPM detector evaluation board from Broadcom, which is presented in the Supplementary material Section \ref{sub:sipm_board}. As with the APD, we were able to successfully detect single photons. Compared to the APD, the SiPM has a much shorter effective dead time, since it consists of thousands of independent SPADs and it is very unlikely for two photons to consecutively hit the same pixel. On the other hand, SiPMs are large-area devices, typically on the order of several square millimeters, meaning they have a proportionally larger dark count rate (around 10 MHz at room temperature for our device), which might be problematic.

At this point, it is hard to say which detector is more suitable for our application without more detailed characterization. Both the APD and the SiPM meet our dead time requirements, although the SiPM has much more legroom in this department. We suspect the APD might be a better choice due to a lower dark count rate, but it is unclear how difficult it would be to couple such a tiny chip (area of \SI{0.04}{\milli\meter\squared}) to an optic fiber. If a large area detector is preferred, the SiPM would be a more obvious choice. Another important factor in the comparison is the timing performance (jitter), which we haven't been able to measure yet.

\subsection{Time-to-digital converter (TDC)}
For precise time measurements of signals, we need a device called a time-to-digital converter (TDC), which is essentially a sophisticated digital stopwatch. It represents a vital part of every time-of-flight system. There are many commercial TDC chips available with resolutions in the order of 10 ps, such as the Texas Instruments TDC7200 series, Maxim Integrated MAX35102/35101, AMS AS6500 etc. However, these are standalone chips with tiny footprints that need a host printed circuit board with all necessary infrastructure, like power supplies and a clock source. They are complex, offer a limited number of channels and use complicated data interfaces to transmit timestamps to an external processor, which needs to be programmed to capture the data before sending it to a host PC.

A more elegant solution for quick prototyping is to implement the TDC on a field programmable gate array (FPGA) platform, which can host an arbitrary number of TDC channels. In the case of modern system on chips (SoCs), the FPGA has integrated CPU cores which can easily interface the TDC channels, process the timestamps and conveniently send them to a host computer. For our demo application, we decided to use an open-source 2-channel time-to-digital converter (TDC) system \cite{Adamic_TDC, tdc_github} implemented on a Red Pitaya STEMlab 125-14 board \cite{redpitaya}, which features a Xilinx Zynq 7010 FPGA/SoC. The details of the implementation and a basic performance characterization of the TDC are presented in Supplementary Section \ref{sub:redpitaya_tdc}.

The selected TDC has an advertised dead time of 14 ns, which enables capturing up to 70 MS/s and offers a theoretical maximum timing resolution of 11 ps per channel. With our setup, we measured an inter-channel resolution of around 40 ps, which could be brought further down to 20 ps by using better cables and signal integrity measures. This seems more than good enough for our target application.

\subsection{Full Time-of-Flight System}
\begin{figure}[hbtp]
    \centering
    \includegraphics[width=0.9\textwidth]{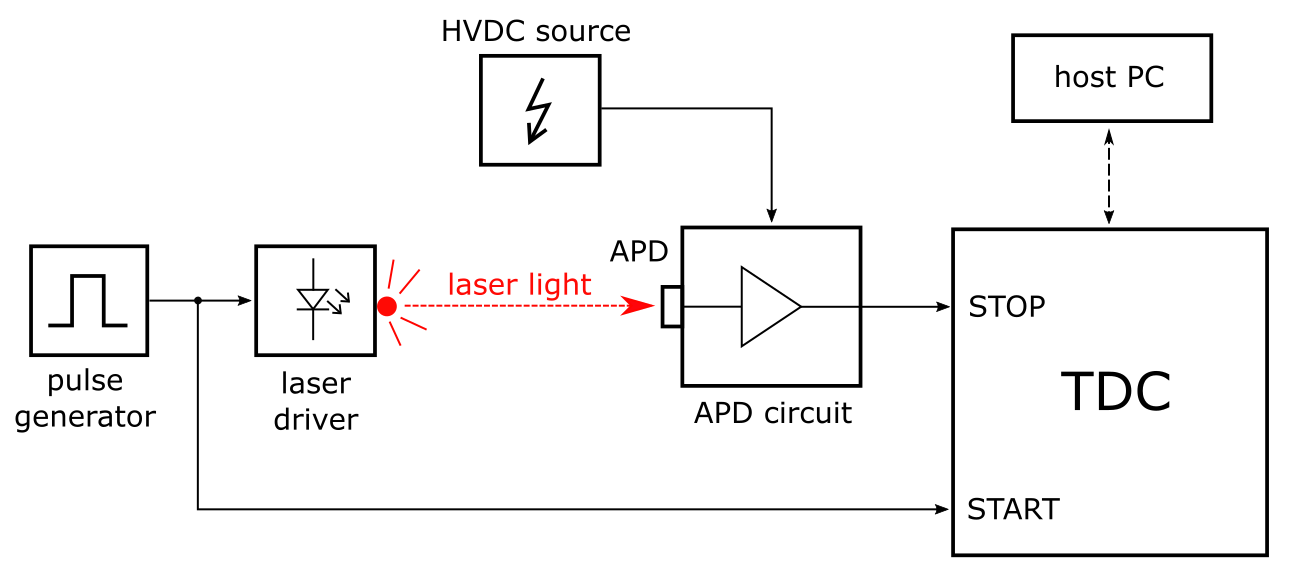}
    \caption{Optical time-of-flight demo system schematic, consisting of a pulsed laser source, a single photon APD detector, and a two-channel TDC. The high voltage power supply provides the bias for the APD, and the TDC is interfaced with a host PC. Both the pulse generator and the TDC are hosted by the Red Pitaya board.}
    \label{fig:tof_schematic}
\end{figure}

Finally, we combined the built laser source, the APD detector and the TDC into a full time of flight (ToF) setup, schematically illustrated in Figure \ref{fig:tof_schematic}. We implemented a 1 MHz pulse generator on the Red Pitaya board as well, together with the TDC. See Supplementary Section \ref{sub:tof_system} for full details.

Since we didn't manage to get the laser diode to produce short enough light pulses, we were unable to demonstrate a proper TD-fNIRS instrument response function, i.e. a DTOF histogram. However, we did show that the ToF system works by measuring the 1 MHz modulation of the laser source (see supplementary material).

We are actively working on getting our laser diode to gain-switch, which will allow us to characterize the timing performance of the system more precisely. Such a performance characterization would include measuring the total system time resolution by combining the temporal jitter of the laser source, the APD detector and the TDC. In addition to the DTOF histogram, we would also see a constant background due to the dark counts of the detector. It is imperative to verify that the signal-to-noise ratio (SNR) of the ToF measurement is high enough, i.e. the histogram rises well above the dark background. If the SNR is too low, a different model or a more expensive SPAD must replace the APD.

Another important characterization step is to measure the histogram sampling rate of the system, which is related to the maximum photon count rate of the laser-detector pair. Ideally, this should be as high as possible, above 10 Msps. In our prototype, we used a 1 MHz modulation for the laser due to jittery jumper wires; this was necessary to preserve signal integrity. Making the setup high-frequency-friendly will enable increasing the modulation frequency. Ultimately, the count rate of the APD active quenching circuit (about 20 MHz) and the TDC channels timestamp rate (70 MHz) place limits on the achievable sampling rate.

One of the main challenges of the presented setup is its sensitivity to temperature fluctuations. The laser diode's lasing threshold is coupled to the ambient temperature, causing the light pulse's delay and width to drift when the temperature changes. This sensitivity to temperature could ruin the timing performance of our gain-switching pulsed laser diode. Similarly, the APD's breakdown voltage is highly temperature-dependent. It therefore might be important to control the temperature of these components. While the TDC can be calibrated for temperature changes on the fly, the laser diode and the APD might need their own PID thermal control for temperature stabilization. As for the APD, it might be worth cooling it down as well to minimize dark counts.

The presented design offers a potentially affordable solution for quickly implementing an fNIRS prototype. The biggest investments are the FPGA board and the HVDC source for biasing the detectors -- amounting to about \$1,000 USD -- but they are common to all channels, so the cost per channel goes down as more channels are added. Our pulsed laser source costs around 100 USD (PCB + laser diode), and similarly for the detector circuit (the exact price varies depending on the chosen APD vs SiPM technology). So if we envision a simple 10-channel prototype, with a single laser source surrounded by 10 detectors and read by a 10-channel TDC, we are looking at roughly $\sim$\$2.5 k for the electronics parts. Adding thermal management and optomechanical components might increase that a bit further, but not drastically.

\section{Code availability}
Our project is open source at \url{https://github.com/neural-imagery}.  We share code repositories for our machine learning and tomography experiments, as well as our experimental protocols and data collection software.

\section{Acknowledgements}
We'd like to sincerely thank TechEN for loaning us a CW headset, which was incredibly helpful in understanding the fNIRS signal. We'd also like to thank David Wong-Campos, Simon Arridge, Ted Huppert, and David Boas for helpful discussions, and Simon Arridge for providing time-domain reconstruction code in Toast++. The hardware team would also like to thank Simon Tartakovsky for his advice and his feedback on the paper.

We're especially grateful for Schmidt Futures, the Moth Fund, HackGrants, and Misha Gerovitch, who provided funding for this project.

\nocite{*}
\printbibliography 

\pagebreak

\appendix 
\renewcommand{\thesection}{S\arabic{section}}

{\Large Supplementary Information}
\label{sec:supplementary}
\addcontentsline{toc}{section}{Supplementary Information}
\section{Data quality}
\label{sec:data_quality}

\renewcommand{\thefigure}{S\arabic{figure}}
\setcounter{figure}{0}

We’ve demonstrated the possibility of reconstructing images with fNIRS, but actually doing so and collecting a full dataset comes with its challenges. We made an initial attempt to collect data over two weeks, and we share some of the difficulties encountered and ways to overcome them. 

We worked with a continuous-wave TechEn headset with four sources and eight detectors. Each source emitted two wavelengths of near-infrared light (735 nm and 850 nm), staggered at different frequencies around 6 kHz to differentiate the sources. Data was sampled at 10 Hz.

While recording fNIRS, it wasn’t immediately obvious what constituted a good signal. High photon counts generally correspond to high SNR, but this can be confounded by ambient lighting or IR light that has leaked or bounced off the hair, without transmitting through the head. With our particular headset, we saw a non-stationary oscillation around 0.5 Hz when the detectors weren’t in contact with any sample, which we attributed to a quirk of the hardware.

The first thing we learned was that the signal was extremely sensitive to contact with the scalp. The detectors and sources were enclosed by a light pipe, which needed to be flush against the skin --- otherwise, reflected light drowned out the signal. The headset assumed a specific curvature, and keeping all 12 detectors and sources flush against the head required applying a surprising amount of pressure.

\begin{figure}[hbtp]
    \centering\includegraphics[width=0.8\linewidth]{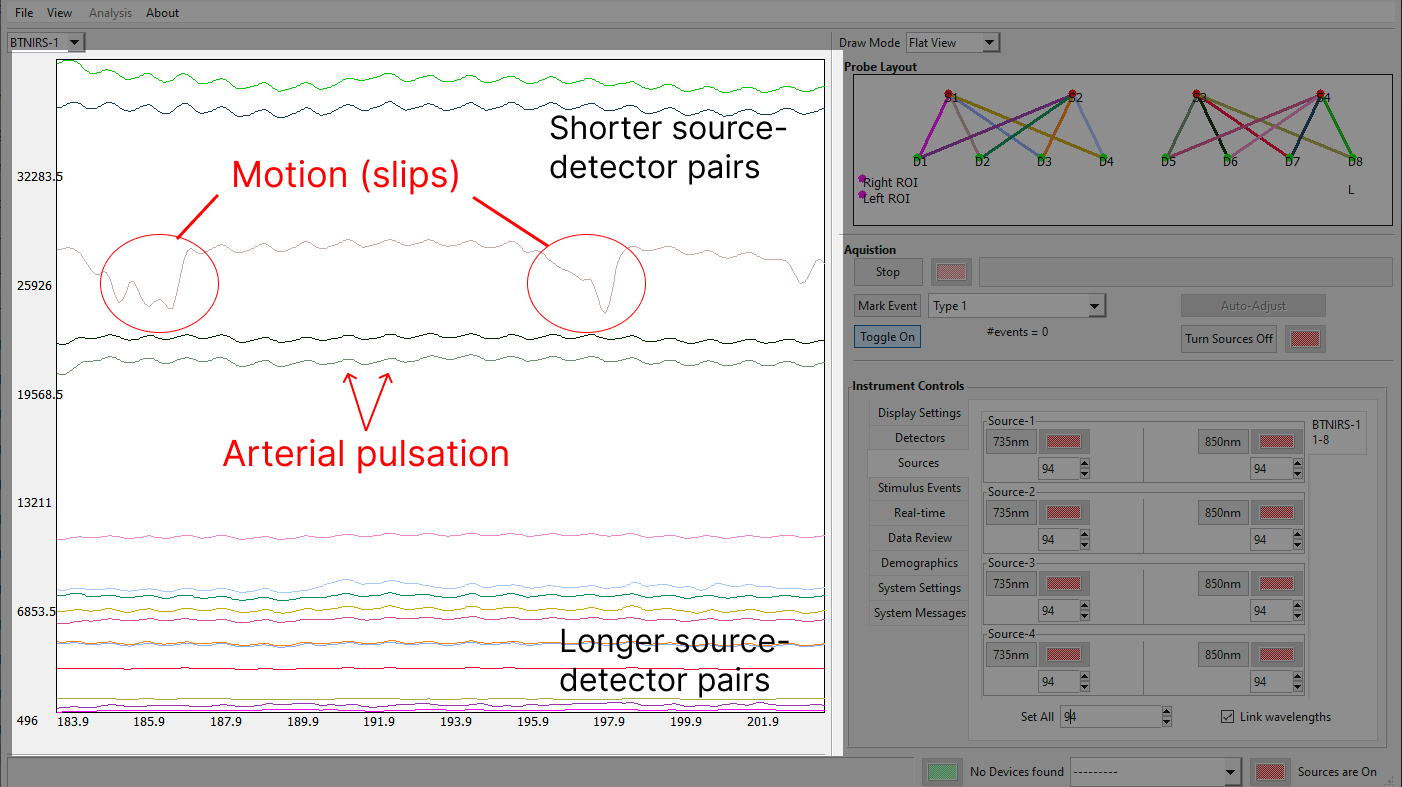}
    \caption{TechEn Dashboard with time-series recording of 16 channels of raw optical intensity.}
    \label{fig:dashboard}
\end{figure}

A signal with good contact is shown in Figure~\ref{fig:dashboard}. Good quality signal shows a clear a periodic pulse in the raw light readings, indicating arterial pulsation. 

We started with a hairless region, the forehead, and sought out a physiologically relevant signal. We asked our participant to hold their breath for 20 seconds and tried to detect a decrease in oxygen levels, as per \cite{oxygenbrain}. We were unable to observe a consistent decrease in oxygenation during voluntary breath-holding (after filtering and applying Beer-Lambert’s law).

\begin{figure}[hbtp]
    \centering\includegraphics[width=0.75\linewidth]{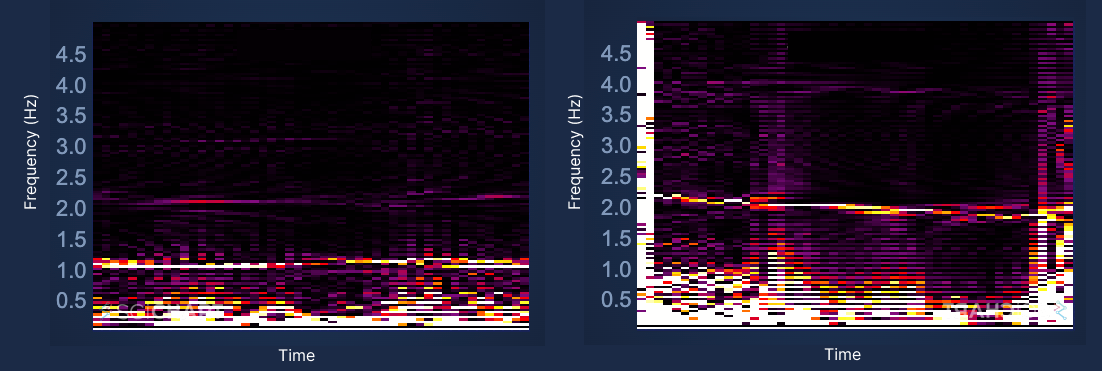}
   \caption{Spectrograms (time-frequency maps) for a single source-detector pair before (left) and after (right) exercise.}
    \label{fig:spectrogram}
\end{figure}

Seeing nothing, we had our participant run up and down the stairs for 30 seconds. The arterial pulsation went from 1Hz at rest to 2Hz after exercise, and we watched the frequency peak drop slowly as the participant's heart rate decreased after the run. The signal was extremely obvious when plotted as a spectrogram (Figure \ref{fig:spectrogram}).

\section{Retinotopy}
\label{sec:retinotopy}
As a next step towards fNIRS image reconstruction, we tried to classify the spatial region of visual stimuli from visual cortex activation. Retinotopy, or the spatial organization of retinal data on the cortex, is usually studied with fMRI or PET  \cite{hoffmann2009}. In 2005, bilateral activation of the visual cortex was shown using fNIRS and validated against simultaneous fMRI \cite{pmid:zhang}. Since then, high-density systems have been successful in constructing a retinotopic map of checkerboard wedges at steps of 10° of rotation, first in a fiber-based system \cite{pmid:egg} and recently in a wearable fNIRS \cite{lumo}.

Our system had far fewer channels, and we performed our experiments in a non-laboratory setting. We were able to see good discriminability between left and right visual stimulation using 4 channels on each side of the brain.

\subsection{Methods}
\textbf{Subject}. CW-fNIRS data were collected from a healthy 23-year-old male subject. 

\textbf{CW-fNIRS acquisition}. The optodes were positioned symmetrically on the region of the scalp over the occipital lobe. This region on the scalp was shaved prior to data collection to improve contact and signal quality.

\begin{figure}[!h]
    \centering
    \includegraphics[width=0.75\linewidth]{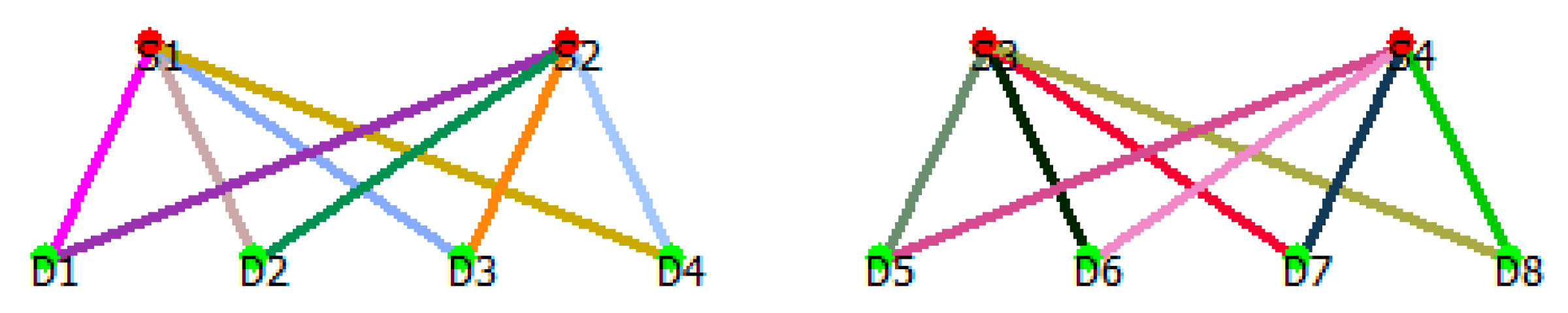}
    \caption{The TechEn headset's 16 data channels, made up of 4 sources (S1 to S4) and 8 detectors (D1 to D8).}
    \label{fig:sources_detectors}
\end{figure}

The sources and detectors were arranged and paired as seen in Figure \ref{fig:sources_detectors}, to create 16 channels in each wavelength, for a total of 32 data channels across the two wavelengths.

\begin{figure}[hbtp]
    \centering\includegraphics[width=0.75\linewidth]{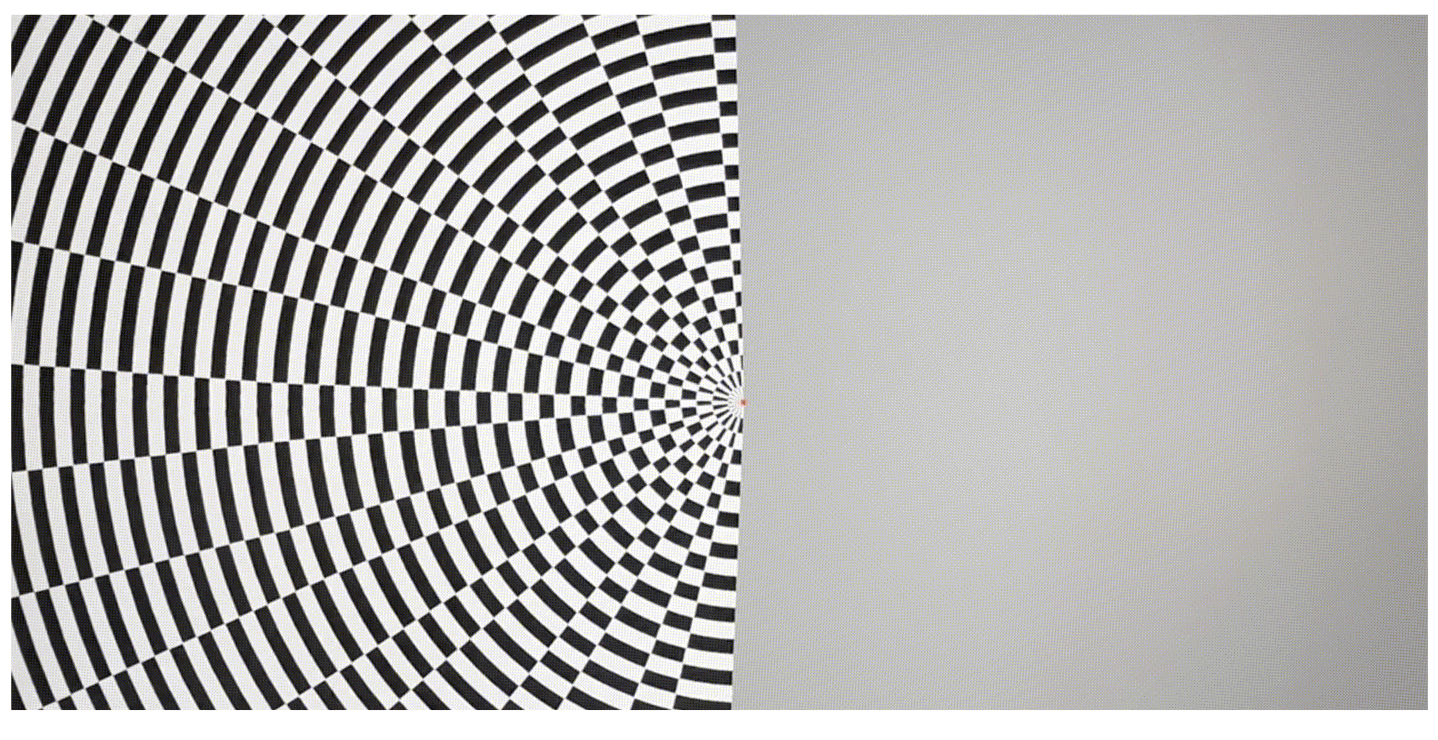}
   \caption{Visual stimuli for the retinotopy experiment, "Left" condition.}
    \label{fig:checkerboard}
\end{figure}

\textbf{Visual stimulation}. The stimulus pattern was based on a circular checkerboard consisting of black and white checks (see Figure \ref{fig:checkerboard}). The subject fixated on a central, red dot during the experiment. For a coarse (binary) retinotopic classification, one half of the visual cortex was stimulated with the flashing checkerboard pattern while the other half remained gray \cite{hoffmann2009}.

The visual experimental paradigm was implemented using the PsychoPy package and presented on a laptop computer. Each trial consisted of a 5-second stimulation period followed by a 15-second rest period. The subject was instructed to look at a fixation point in the middle of the screen for the entire experiment. During the stimulation periods, a black-and-white checkerboard pattern filled out one half of the screen (left or right). The checkerboard flickered at a rate of 5 Hz. A static background was presented during the rest periods. We performed 8 runs for a total of 80 trials for each stimulus condition.

\subsection{Results}
Raw fNIRS data were processed using the Homer3 MATLAB application \cite{HOMER}. The standard Homer processing pipeline was used with default parameters. We discarded data from the two most lateral sources (i.e. the sources that were furthest away from the midline) after visual inspection.

\begin{figure}[hbtp]
    \centering
    \includegraphics[width=0.75\linewidth]{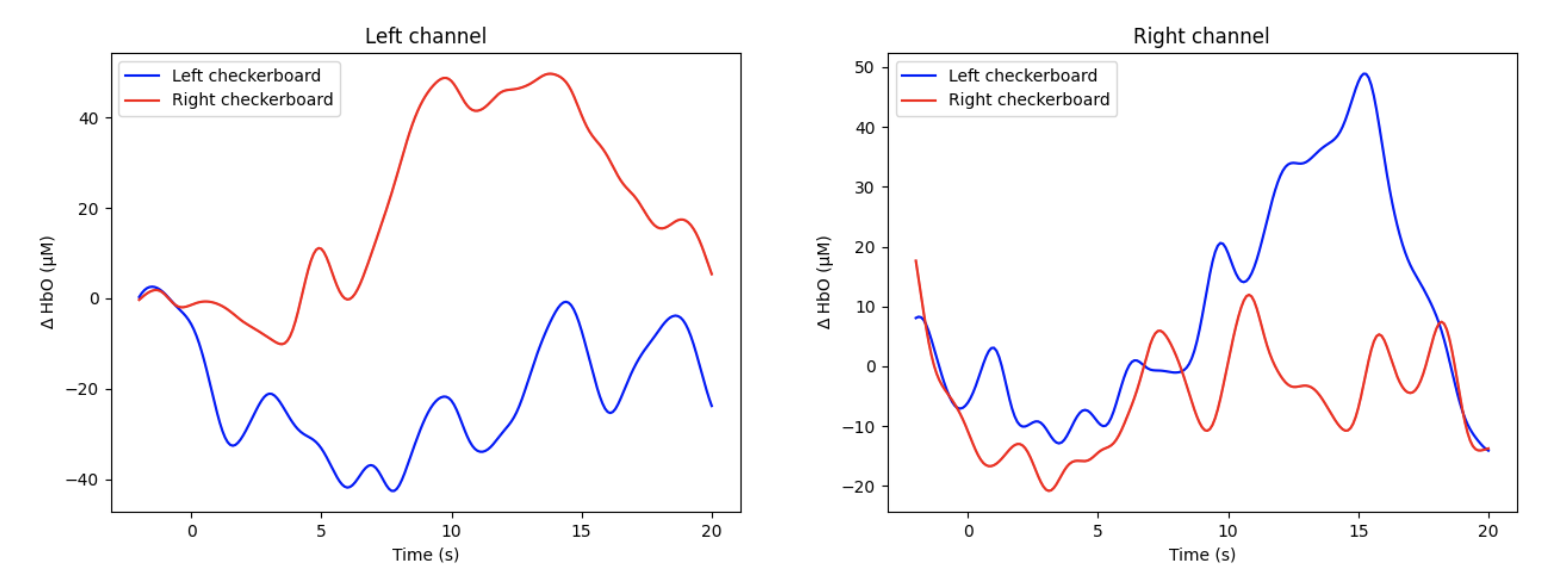}
   \caption{Average oxyhaemoglobin concentration for left and right channels during visual stimulation.}
    \label{fig:avg_hbo}
\end{figure}

Figure \ref{fig:avg_hbo} shows the average oxyhaemoglobin concentration (HbO) from source-detector pairs from the left and right side of the head for each stimulus condition. There is a clear difference between the two conditions, and we see the expected haemodynamic response: when the flickering checkerboard was presented in the right visual field, there is an increase in the HbO measured from source-detector pairs on the left side of the head; the opposite pattern can be observed when visual stimuli was presented in the left visual field.

A machine learning classifier trained on a balanced set of 96 trials of the data and tested on a balanced holdout set 24 trials was able to predict the stimulus condition with an accuracy of 75\% in the holdout set. We used a random forest model, with inputs of the mean of 5-second chunks of data, from 0-5 s to 15-20 s post start of event. The model and data are available at \url{https://github.com/neural-imagery/signal-classifier}.

\section{Data collection dashboard}
\label{sec:dashboard}
It was extremely important to visualize signal features in real-time and adjust the headset until it was recording clean data. We built a dashboard that shows a spectrogram of each channel (see Figure \ref{fig:spectrogram} for a snapshot). The dashboard is at \url{https://github.com/neural-imagery/neural-imagery-app}.

\section{Time domain prototype}\label{sec:td_prototype}
Here, we present the implementation details of the components used in our TD-fNIRS prototype. This includes a fast pulsed laser driver, APD and SiPM single photon detector circuits and a two-channel TDC system on a Red Pitaya FPGA board.

\subsection{Pulsed laser driver}\label{sub:laser_driver}
\begin{figure}[hbtp]
    \centering
    \includegraphics[width=0.8\textwidth]{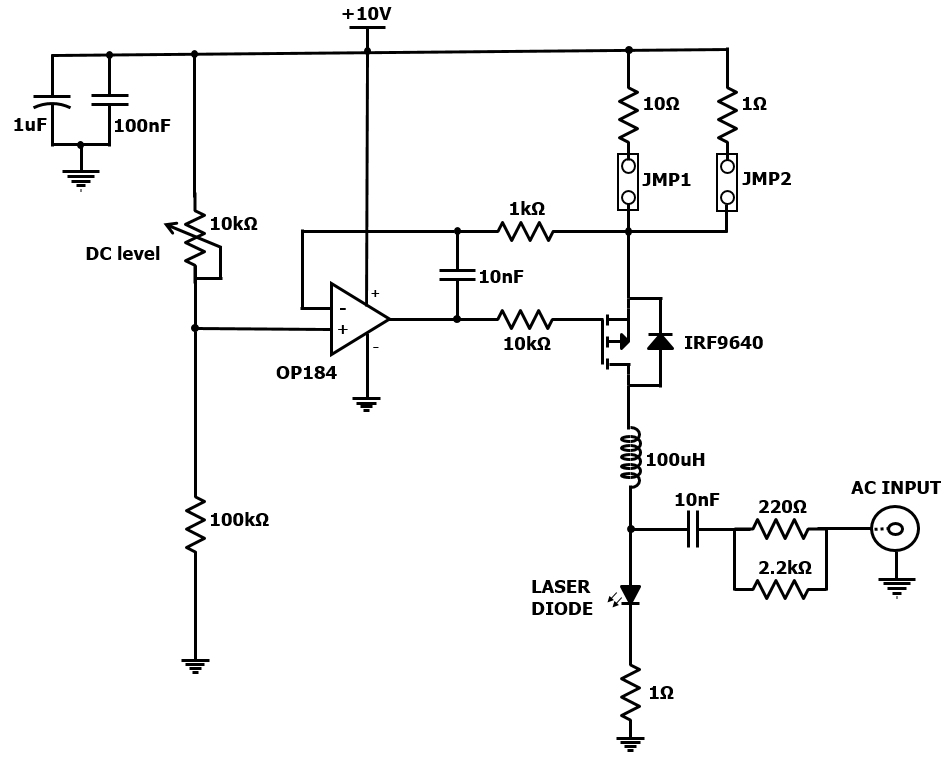}
    \caption{Schematic of the laser driver circuit. The \SI{10}{k\ohm} potentiometer sets the DC current through the laser diode, while the AC (square wave) component is brought in from outside. The \SI{10}{nF} capacitor on the output of the operational amplifier is needed for stability.}
    \label{fig:laser_driver}
\end{figure}

To implement laser gain switching, we developed and tested the circuit in Figure \ref{fig:laser_driver}, first on a breadboard and then on a printed circuit board (PCB). It's a DC current source, tunable from \SIrange{0}{80}{\milli\ampere} or \SIrange{0}{800}{\milli\ampere} (depending on the jumper setting), combined with an LC bias tee that superimposes an AC input signal on top of the DC bias. The AC signal comes from an external square wave generator, which is the FPGA in our case.

\begin{figure}
    \begin{center}
        \begin{subfigure}{0.45\textwidth}
            \includegraphics[width=\textwidth]{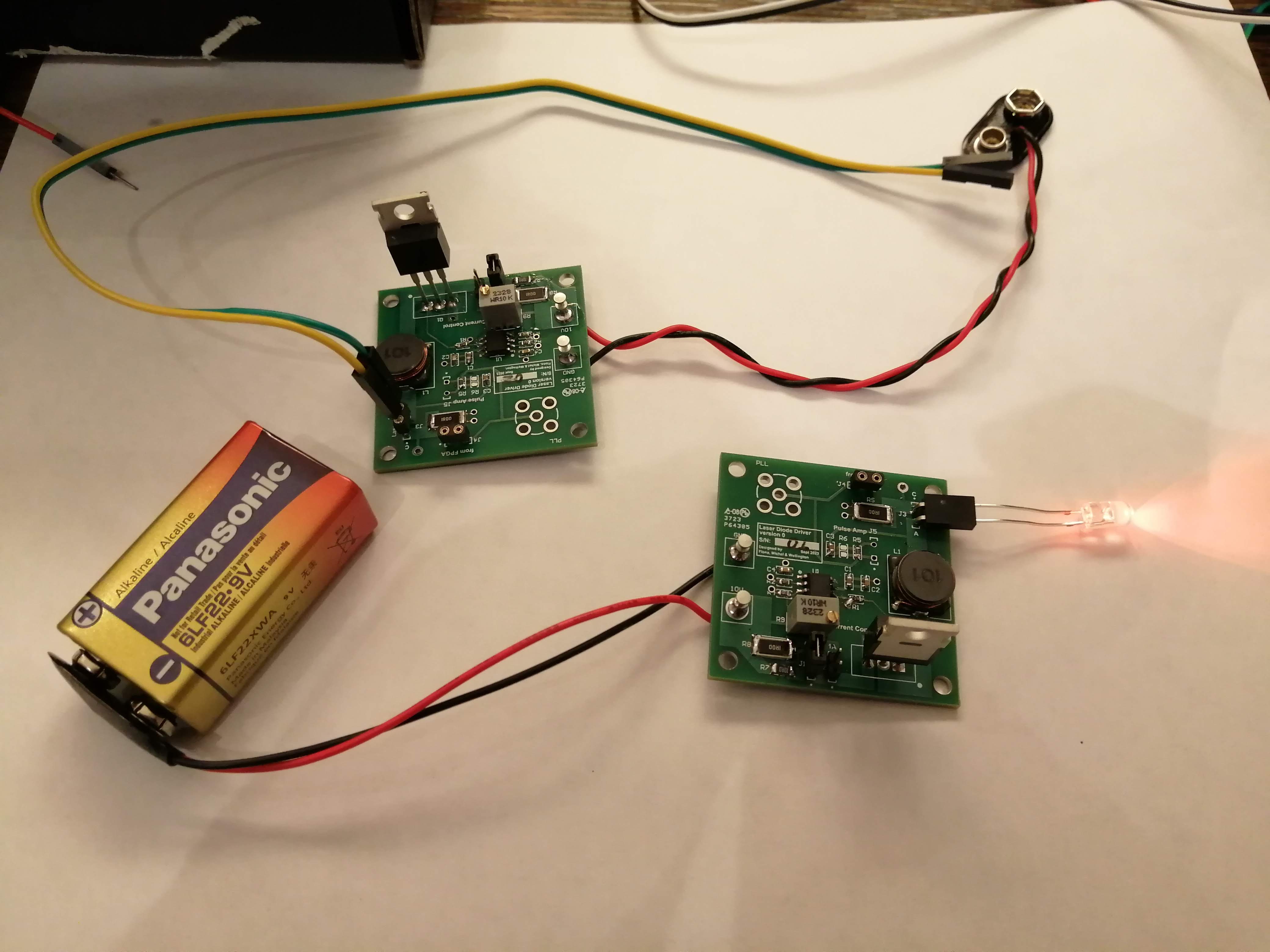}
        \end{subfigure}
        \begin{subfigure}{0.45\textwidth}
            \includegraphics[width=\textwidth]{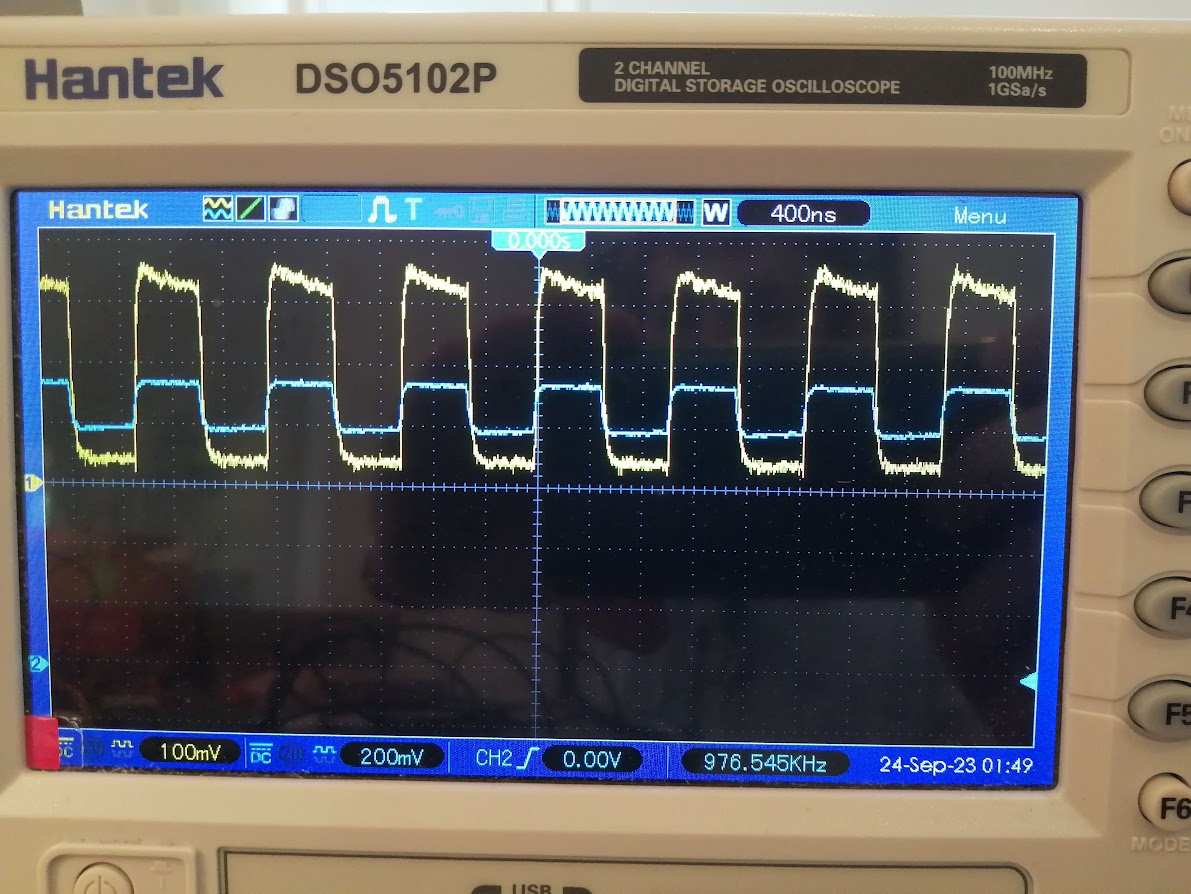}
        \end{subfigure}
    \end{center}
    \caption{Left: Fully assembled laser driver PCBs with a test LED. Right: Blue is the laser diode current, modulated with 15 mA at 1 MHz, and yellow is the APD response. The laser is kept at threshold, so the small modulation has a large effect on its light output.}
    \label{fig:laser_output}
\end{figure}

The circuit was tested with an HL6738MG 690 nm laser diode from ThorLabs with a threshold current of around 65 mA. We used an avalanche photodiode (APD) in linear mode to monitor its light output. The laser was modulated with a 1 MHz square wave of amplitudes 15 and 1.5 mA, respectively. Waveforms of the laser current and the produced light output are shown in Figure \ref{fig:laser_output}.

Unfortunately, no gain switching was observed at 15 mA nor 1.5 mA modulation and the light output was always 500 ns wide as we varied the DC bias current. We suspect we might need to further increase the AC frequency and reduce its amplitude to achieve gain switching. Work is underway to attempt to achieve this.

\subsection{APD active quenching circuit}\label{sub:apd_circuit}
\begin{figure}[b!]
    \centering
    \includegraphics[width=0.6\textwidth]{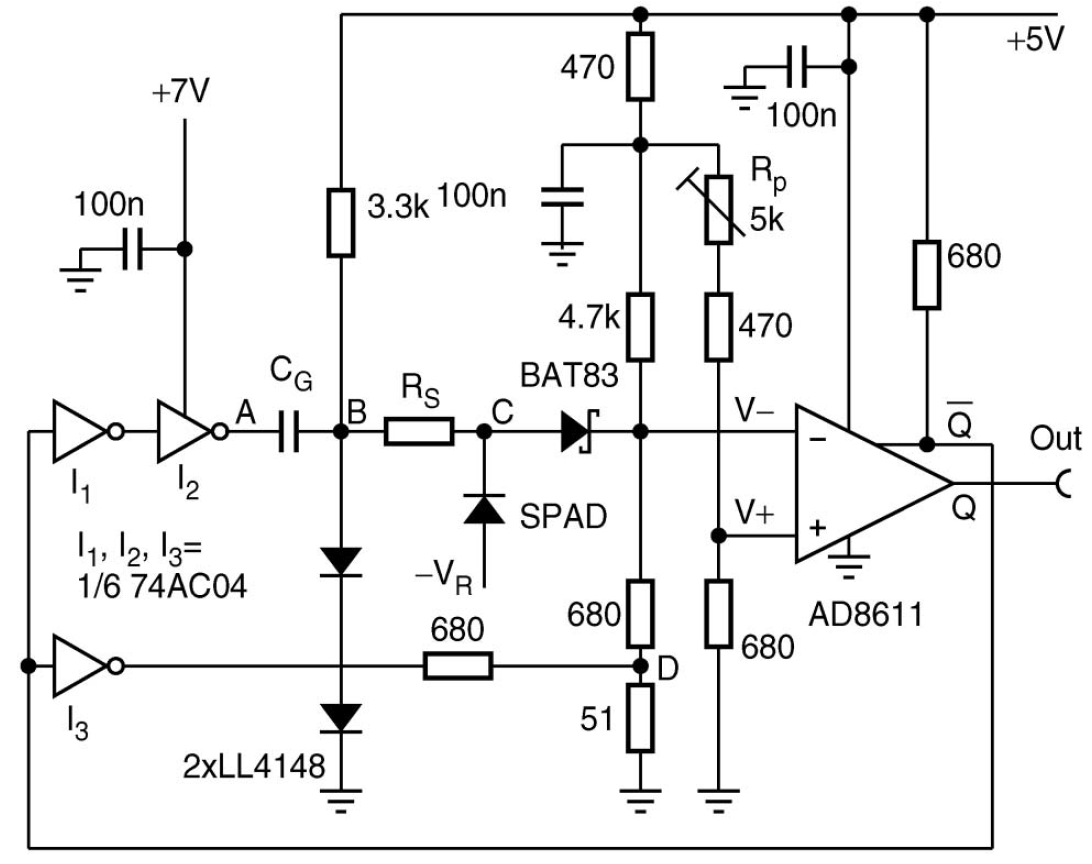}
    \caption{We implemented the APD active quenching circuit proposed by M. Stipčević \cite{Stipcevic_2009}. The figure is from \cite{Stipcevic_2009} and does not include the linear voltage regulators for the +5V and the +7V rails. The APD/SPAD is inserted into the board and -VR is brought in from an external HVDC supply.}
    \label{fig:quenching_circuit}
\end{figure}

We based our APD detector circuit on the design proposed by M. Štipčević \cite{Stipcevic_2009}. Figure \ref{fig:quenching_circuit} shows the circuit diagram.

\begin{figure}[t]
    \begin{center}
        \begin{subfigure}{0.38\textwidth}
            \includegraphics[width=\textwidth]{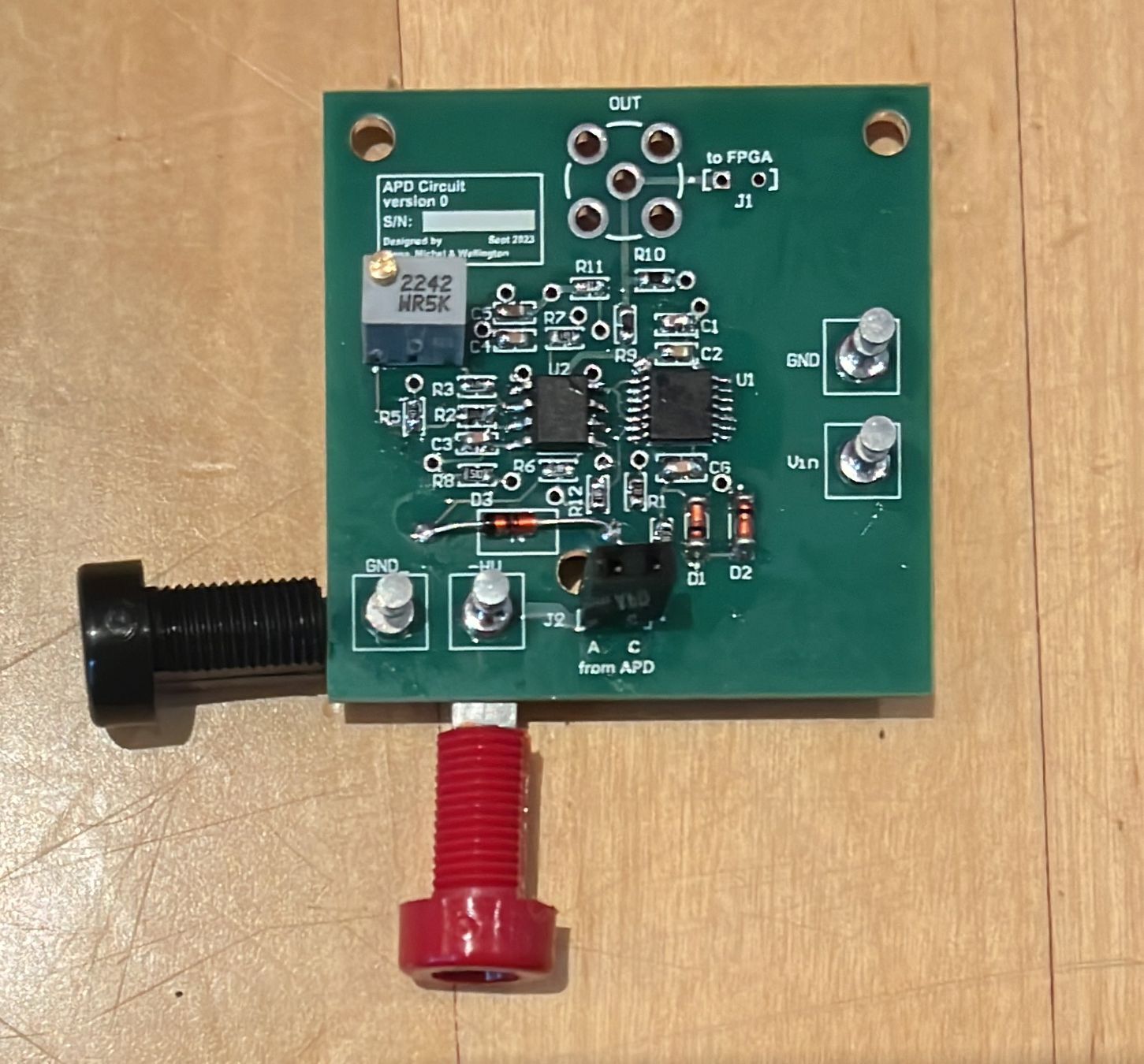}
        \end{subfigure}
        \begin{subfigure}{0.47\textwidth}
            \includegraphics[width=\textwidth]{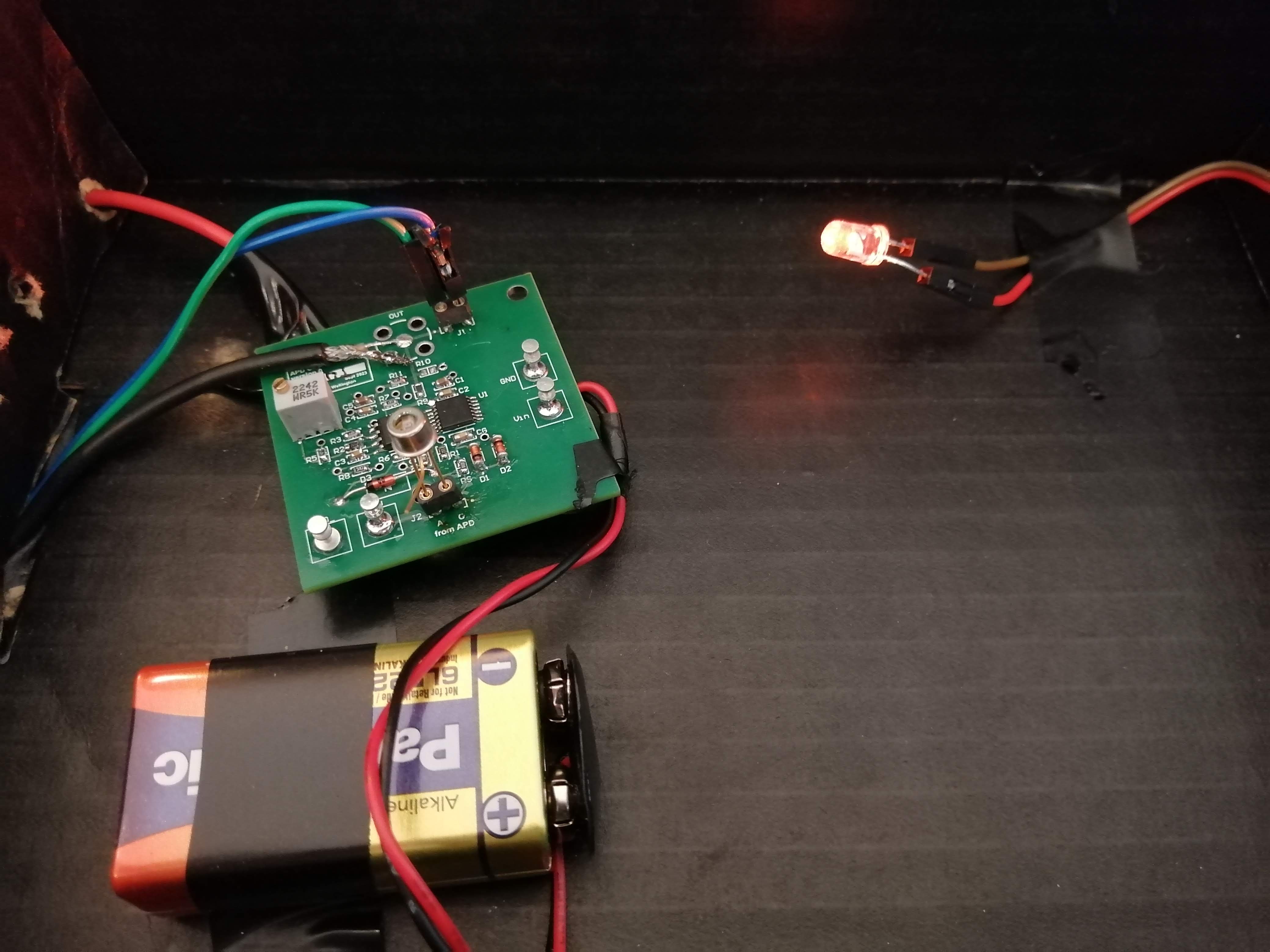}
        \end{subfigure}
    \end{center}
    \caption{Left: assembled PCB of the active quenching circuit. Right: test setup with a Hamamatsu S12023-02 APD and an LED source inside a dark box. The LED was controlled with the laser driver circuit, so we could precisely regulate its light output.}
    \label{fig:quenching_pic}
\end{figure}

The design has a dead time of about 40 ns, allowing for count rates up to about 20 MHz. We coupled the circuit with a Hamamatsu S12023-02 Si APD, which is not optimized for single photon Geiger mode operation but performed very well nonetheless, likely due to successful active quenching. The assembled PCB and a dark box setup for light detection characterization are shown in Figure \ref{fig:quenching_pic}.

\begin{figure}[hbtp]
    \begin{center}
        \begin{subfigure}{0.42\textwidth}
            \includegraphics[width=\textwidth]{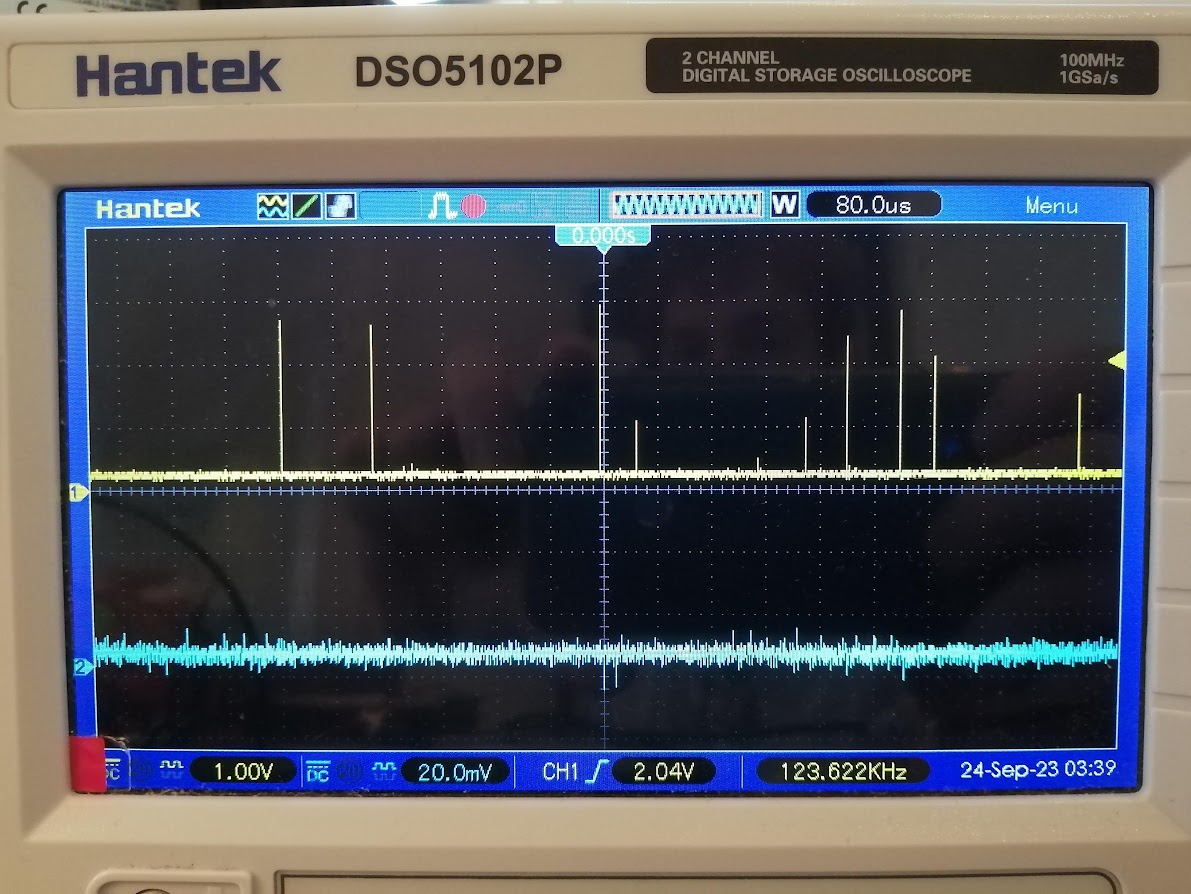}
        \end{subfigure}
        \begin{subfigure}{0.42\textwidth}
            \includegraphics[width=\textwidth]{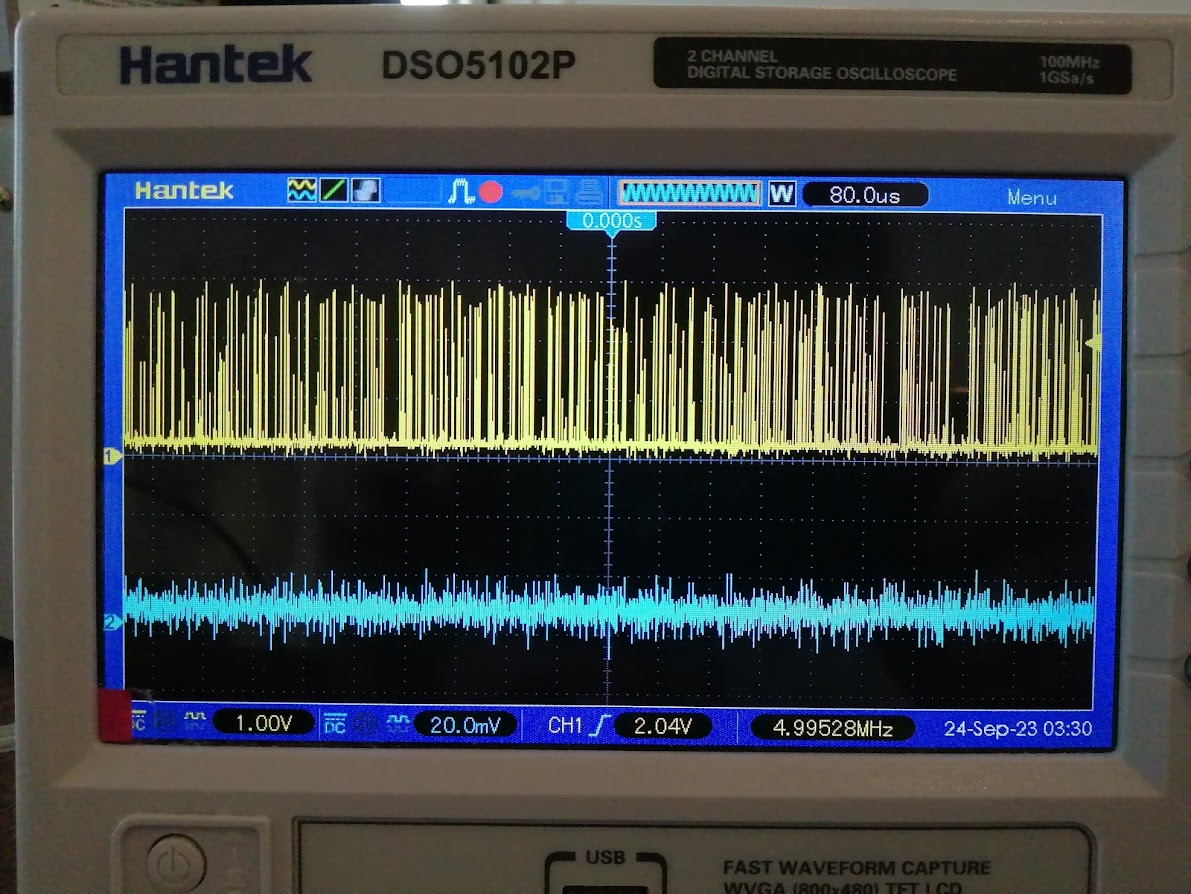}
        \end{subfigure}
    \end{center}
    \caption{APD circuit performance. Yellow is APD digital output, blue is LED current. Left: dark counts. Right: LED current has been very slightly increased, producing a lot of photon detections on the APD. The pulses are a couple of nanoseconds wide, as expected \cite{Stipcevic_2009}.}
    \label{fig:apd}
\end{figure}

The APD was closed inside a dark, sealed box and the digital output of the circuit was observed on the oscilloscope. We increased the high voltage bias on the APD until it reached the Geiger mode of operation at -135 V. This transition is easily observed on the oscilloscope, as dark counts start to appear (see Figure \ref{fig:apd} left). Then, we slowly turned on the LED, producing a lot of photon counts (Figure \ref{fig:apd} right).

\clearpage

\subsection{SiPM evaluation board}\label{sub:sipm_board}
\begin{figure}[p]
    \centering
    \includegraphics[width=0.85\textwidth]{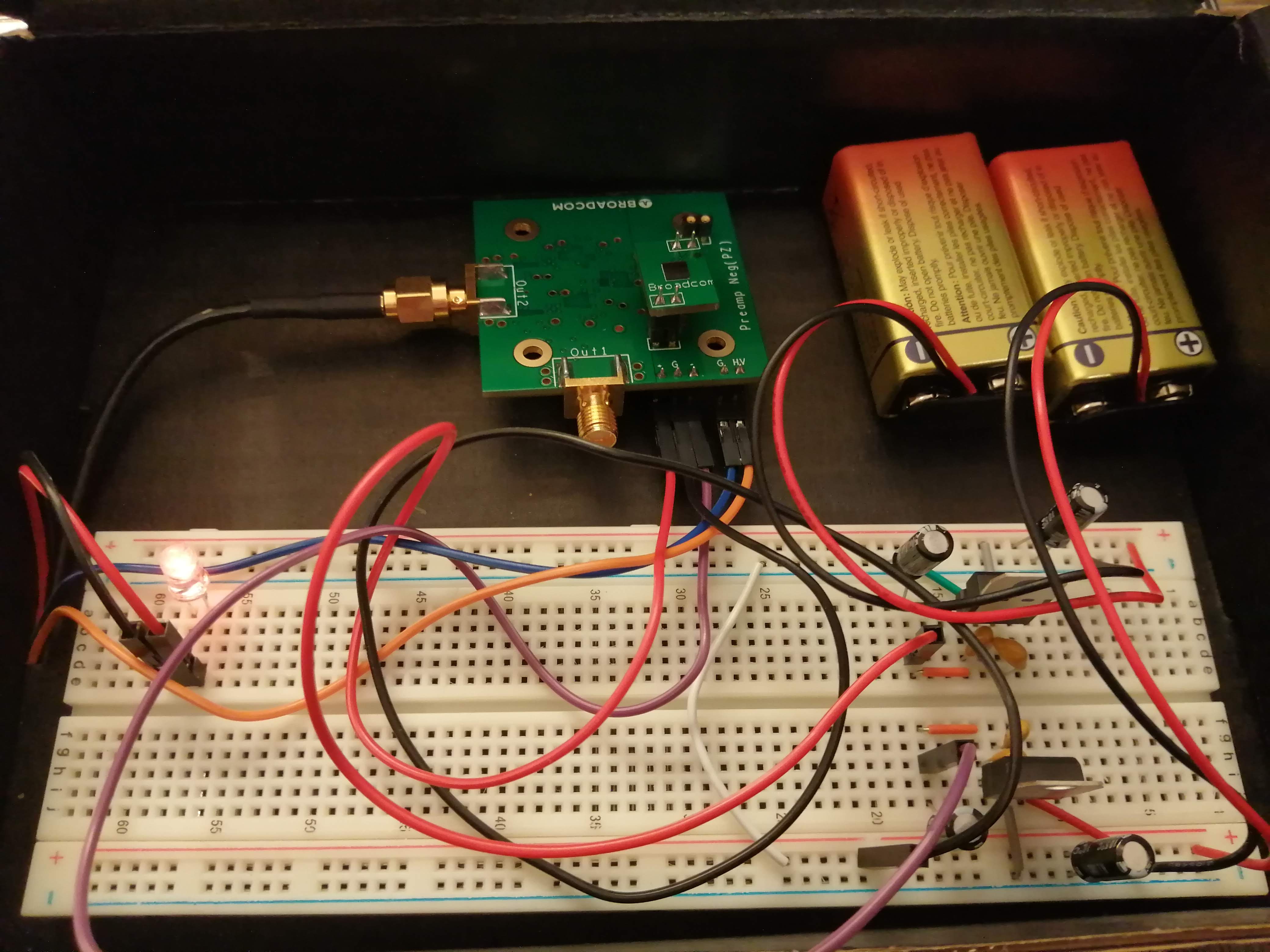}
    \caption{Broadcom’s SiPM and evaluation board in a dark box setup, together with an LED. The breadboard circuit hosts a dual linear 5 V power supply for the evaluation board.}
    \label{fig:sipm}
\end{figure}

\begin{figure}[p]
    \begin{center}
        \begin{subfigure}{0.42\textwidth}
            \includegraphics[width=\textwidth]{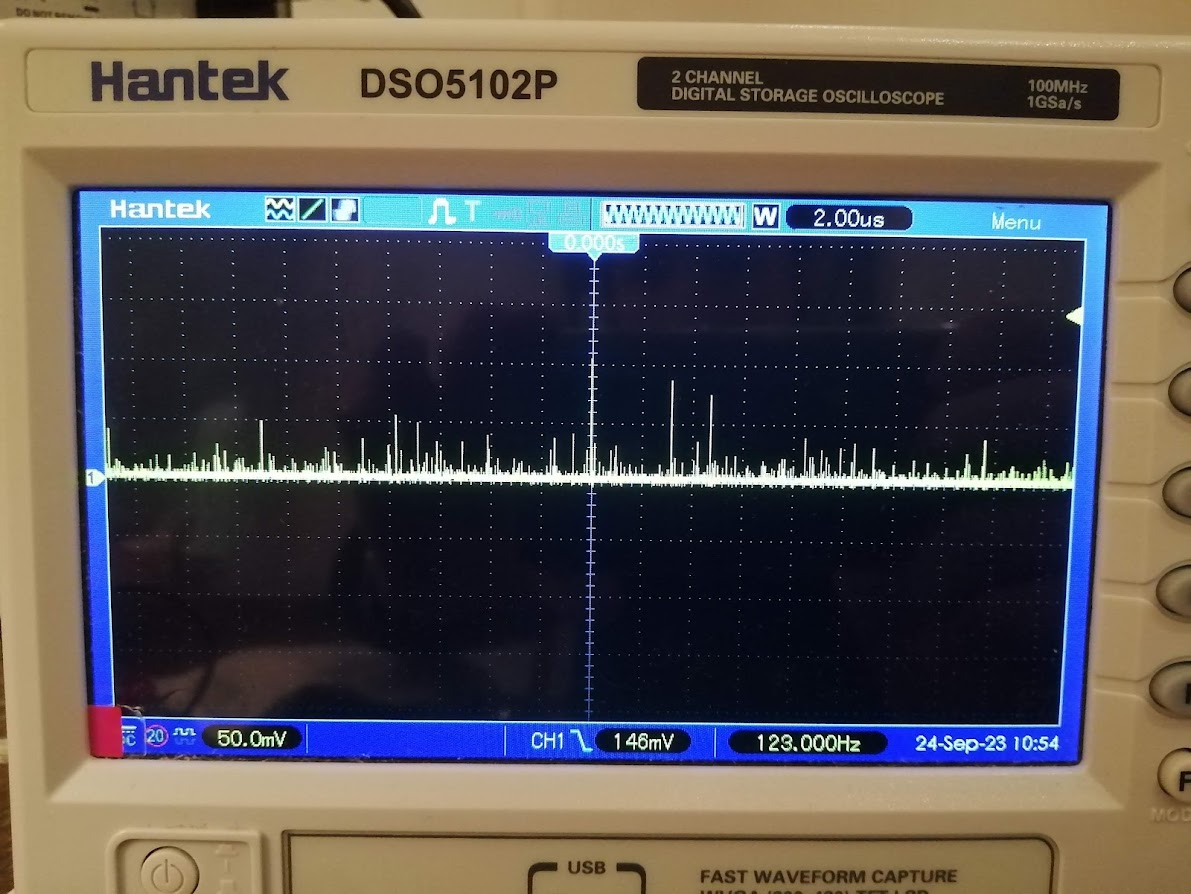}
        \end{subfigure}
        \begin{subfigure}{0.42\textwidth}
            \includegraphics[width=\textwidth]{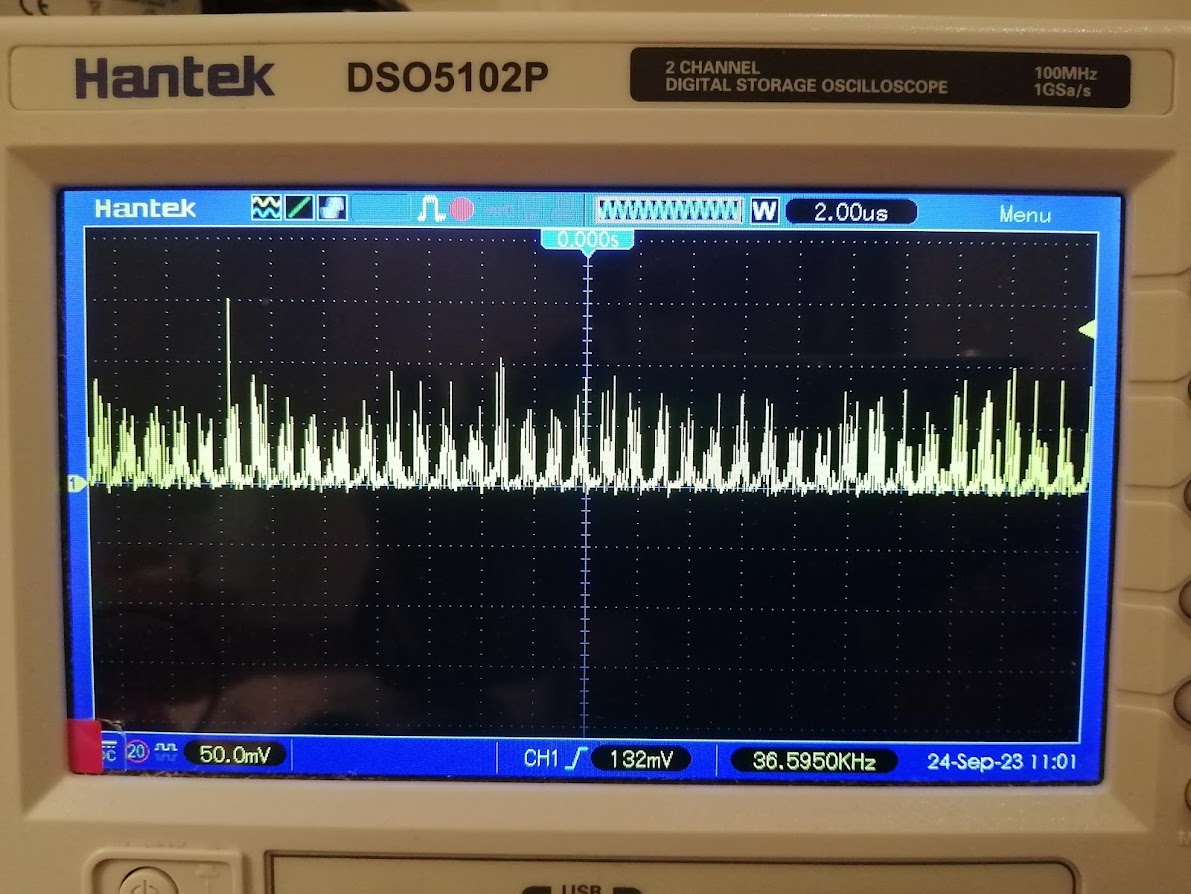}
        \end{subfigure}
    \end{center}
    \caption{Left: dark counts appear as the SiPM device reaches breakdown voltage (Geiger mode). Right: photon pulses dominate as the LED is slowly turned on. Note the clearly visible 1 MHz modulation of the light source at this time scale (2 $\mu$s per division).}
    \label{fig:sipm_data}
\end{figure}

We used the AFBR-S4E001 evaluation board from Broadcom coupled to their AFBR-S4N44C013 SiPM chip for the silicon photomultiplier detector tests. Figure \ref{fig:sipm} shows the dark box setup, which was similar to the one with the APD. The SiPM was biased with +38 V to reach Geiger mode. The results of the light characterization are shown in Figure \ref{fig:sipm_data}; they were made using the fast, analog “OUT2” output of the evaluation board. The LED is first switched off, so only dark counts appear. Then, we turn it on just slightly, causing a lot of photon detections.

The SiPM performed similarly to the APD circuit. However, one major difference compared to the APD is a much shorter (basically negligible) effective dead time, since the SiPM has a lot of independent SPADs that are available to detect an incoming photon. The measured dead time of the SiPM was below the dead time of the TDC, which is 14 ns (not shown here).

\subsection{Red Pitaya TDC}\label{sub:redpitaya_tdc}
\begin{figure}[p]
    \centering
    \includegraphics[width=0.85\textwidth]{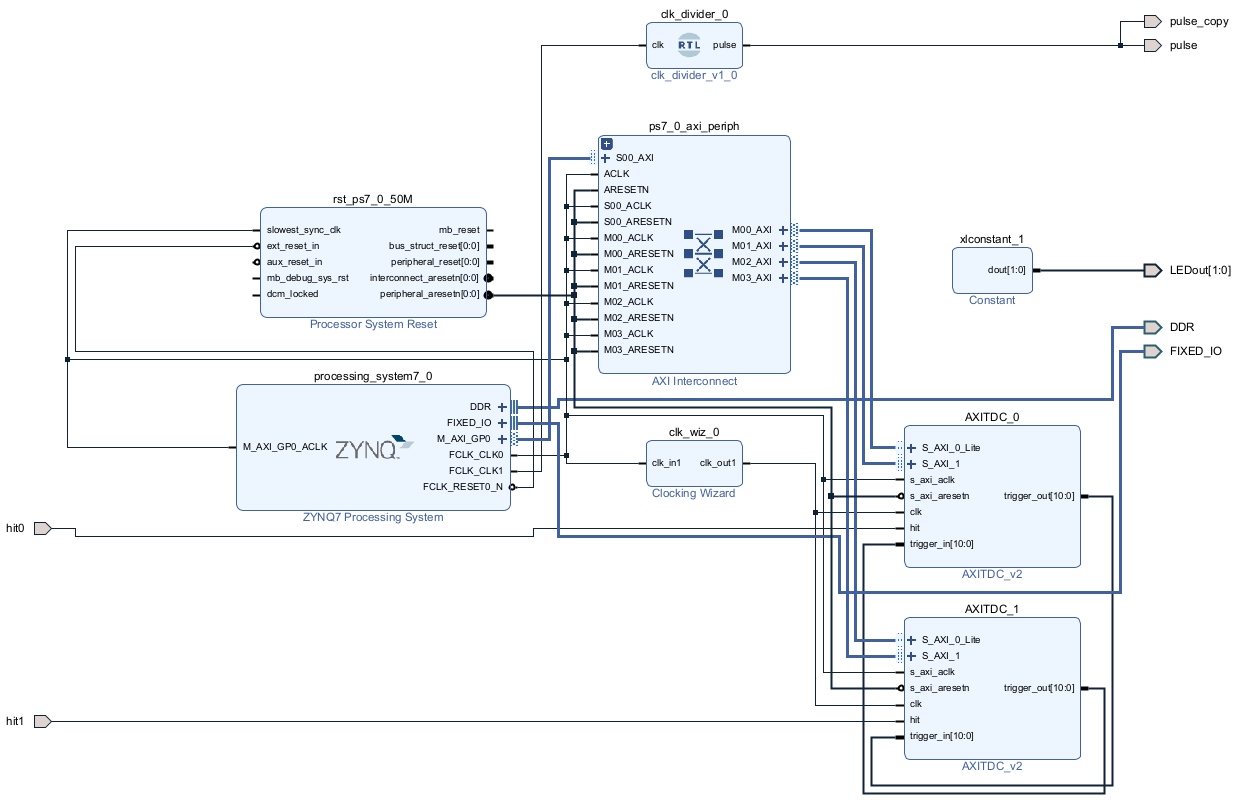}
    \caption{Block design of the implemented TDC system on the Zynq SoC. There are two TDC cores, together with a clock divider that serves as the square wave output. The output frequency is programmable by the Red Pitaya OS.}
    \label{fig:tdc_schematic}
\end{figure}

We implemented a two-channel TDC on a Red Pitaya board using an open-source design by M. Adamič \cite{Adamic_TDC, tdc_github}. The Red Pitaya features a fully programmable Xilinx Zynq 7010 SoC/FPGA, running its own version of the Linux operating system. The TDC design was reimplemented using Xilinx Vivado 2021.2 ML, where we added a square wave generator derived from the processing system clock. We use this square wave output to modulate the laser current and provide the START signal for the TDC. The Block Design of the system implemented in the FPGA is shown in Figure \ref{fig:tdc_schematic}.

\begin{figure}[p]
    \centering
    \includegraphics[width=0.8\textwidth]{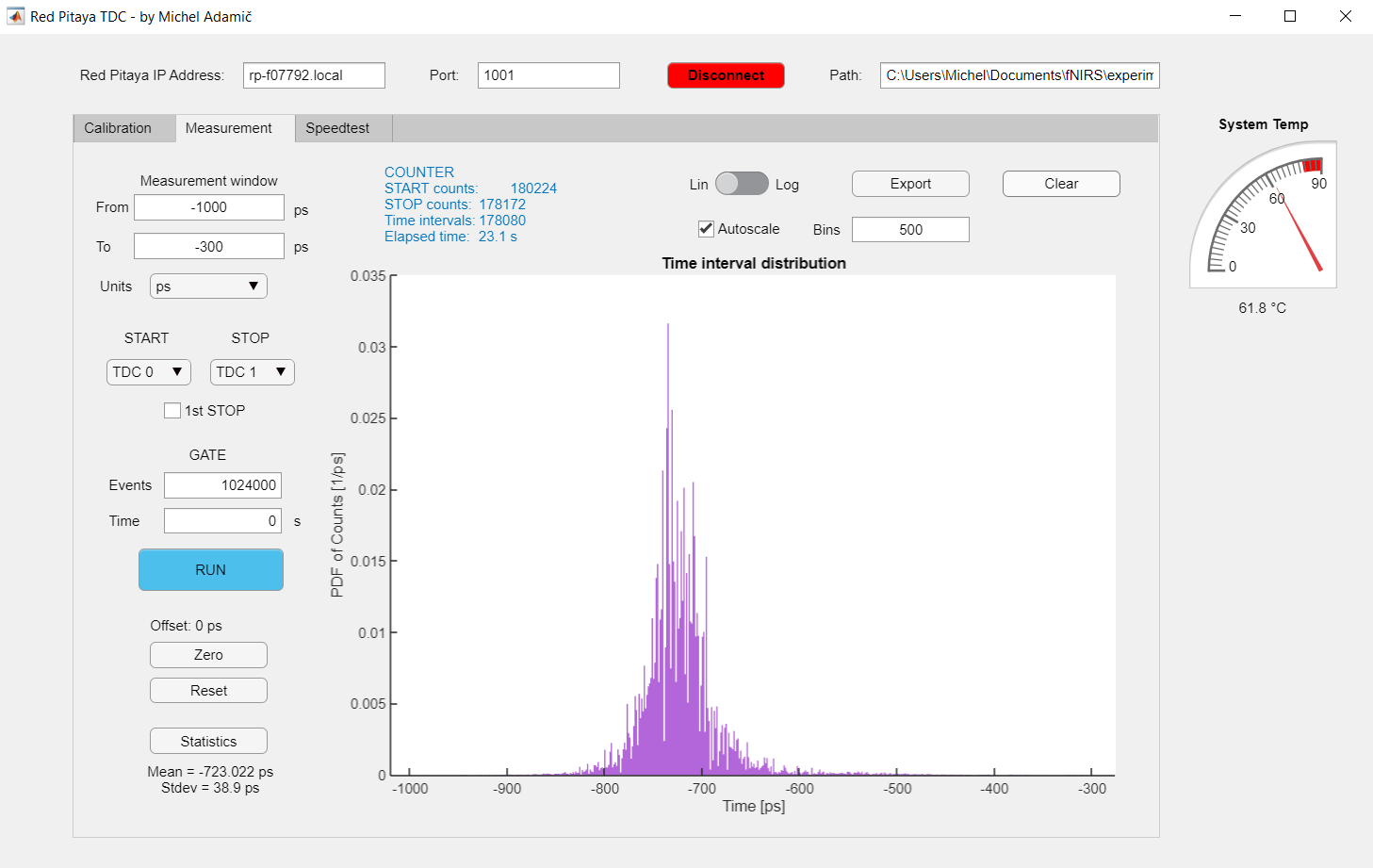}
    \caption{Inter-channel measurement of the same edge yielded a resolution of 39 ps, which is a bit worse than the “advertised” 20 ps or less, but still more than good enough for our needs. We attribute the suboptimal performance of the TDC to the jittery experimental setup, where we used the most basic jumper wires without any proper impedance matching.}
    \label{fig:tdc_data}
\end{figure}

The TDC is interfaced with a MATLAB application running on the host PC. After calibrating the instrument, we performed a basic characterization of the system by connecting the same signal to both TDC channels in order to measure the instrument response function of the TDC itself. The result is shown in Figure \ref{fig:tdc_data}. The inter-channel resolution of the TDC setup was below 40 ps, which is more than sufficient for our needs as we expect much higher jitter from the optical sources and detectors. The resolution was worse than advertised, which can be attributed to a very rudimentary setup with basic jumper wires and total disregard for high speed signal integrity. A proper setup would bring the performance of the TDC closer to 20 ps.

\clearpage

\subsection{ToF system setup}\label{sub:tof_system}
\begin{figure}[ht]
    \centering
    \includegraphics[width=0.8\textwidth]{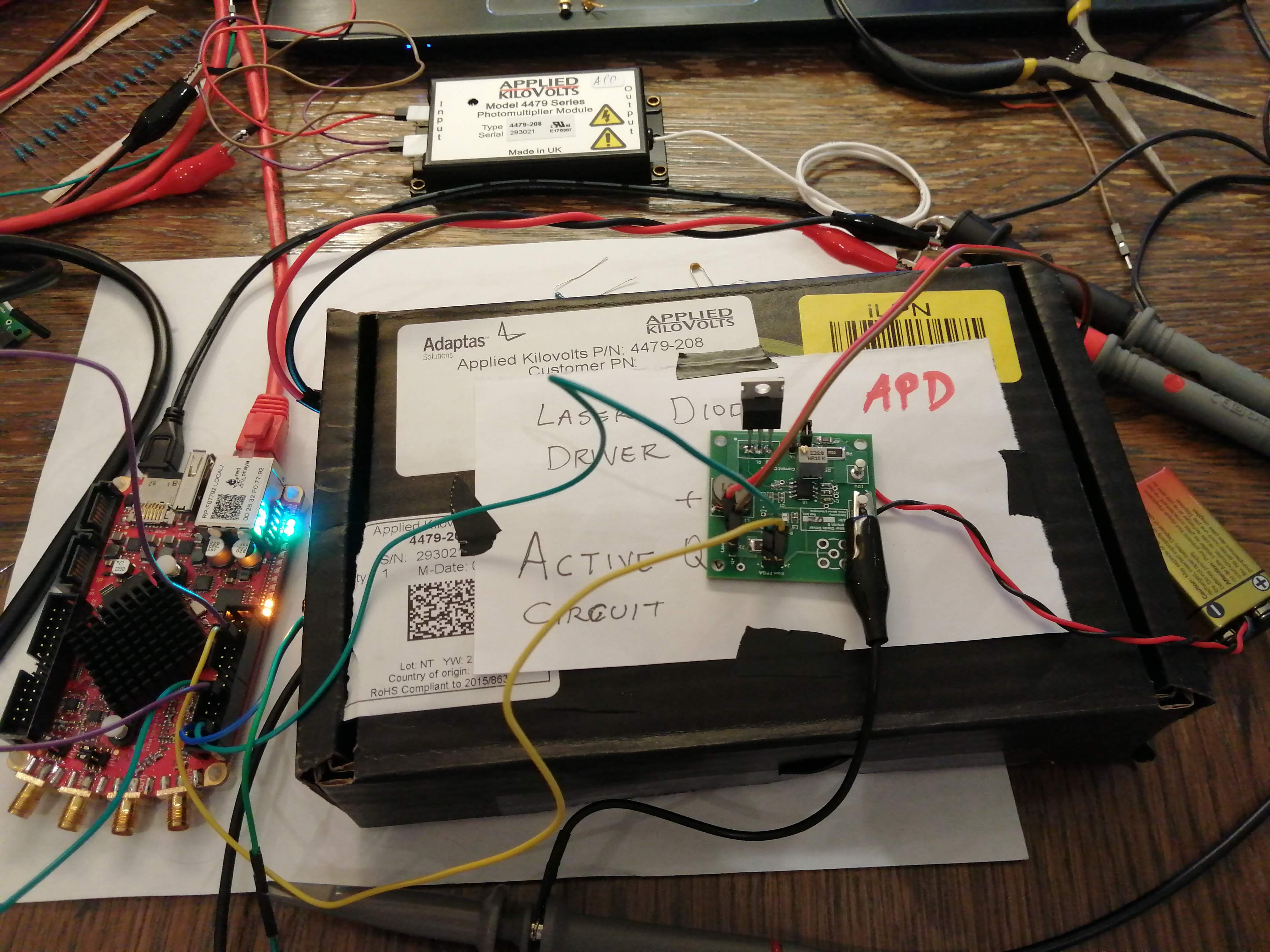}
    \caption{Optical time-of-flight demo system. On the left is the Red Pitaya with the 2-channel TDC system and the square wave generator. On top of the box is the laser driver circuit. Inside the box are the laser diode and the APD circuit presented before. Behind is the Applied Kilovolts 4479-208 HVDC source that provides the negative high voltage bias for the APD.}
    \label{fig:tof}
\end{figure}

Here we present the full ToF system setup we used. It includes the laser driver and a laser diode, the APD detector circuit and the Red Pitaya TDC (see Figure \ref{fig:tof}). A 1 MHz, 3.3 V square wave output of the Red Pitaya serves as the AC modulation signal for the laser driver and also provides the reference (START signal) for the first TDC channel. The 3.3 V digital output of the APD circuit (inside the box) then goes to the second TDC channel to measure the time-of-flight. We are using an Applied Kilovolts 4479 series HVDC source to provide the -135 V bias to the Hamamatsu S12023-02 Si APD.

\begin{figure}[hbtp]
    \centering
    \includegraphics[width=\textwidth]{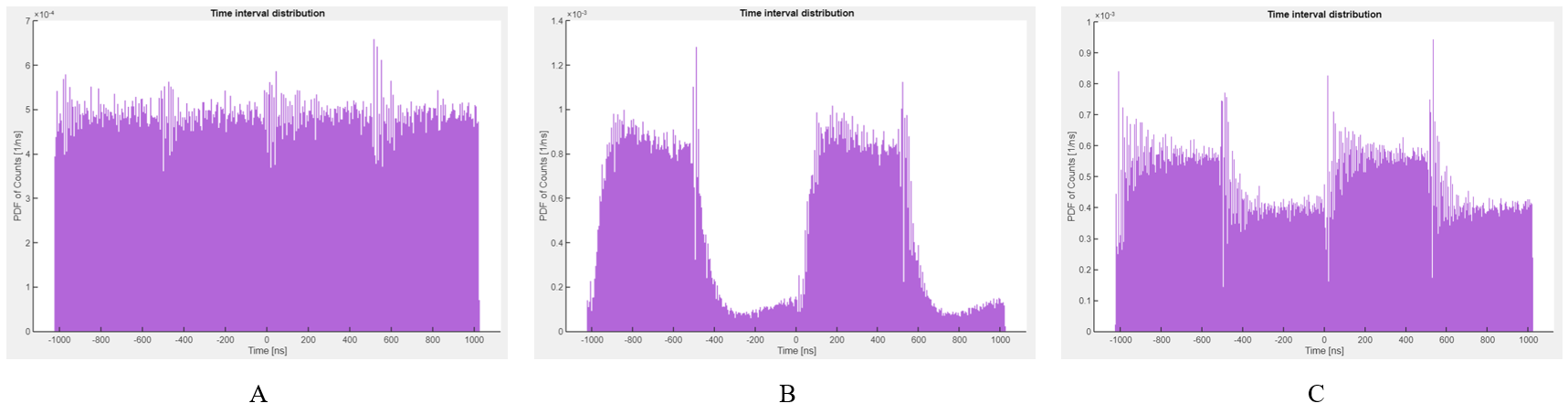}
    \caption{(a) TDC output when the source is turned off, so we are only receiving randomly distributed dark counts. There is however some noise present from the active 1 MHz signal, so distortions at the 500 ns marks can still be seen. (b) TDC output when the source starts to cross the lasing threshold. 1 MHz modulation is clearly visible. (c) Higher light intensity. The 1 MHz modulation is still visible, albeit less clearly. This is because the system (APD + TDC) is starting to saturate in terms of the number of counts it can process. Note: the x-axis time scale spans 2 $\mu$s on all plots.}
    \label{fig:tof_data}
\end{figure}

Since we couldn’t get the laser diode to produce short enough pulses, we were unable to demonstrate a proper TD-fNIRS instrument response function. Nevertheless, we applied a 1 MHz square wave on the laser driver AC input and checked that the TDC can resolve the optical modulation of the light source, using the APD as the light detector. Figure \ref{fig:tof_data} shows the photon arrival time distributions for three distinct cases: when the source is off, slightly crossing threshold, and above threshold. As we can see, the setup can very clearly detect the modulation of the source from the distribution of photon arrival times.

We then repeated the same experiment using the SiPM module instead of the APD circuit. Since the output of the SiPM evaluation board is analog and doesn’t match the LVCMOS33 input levels of the FPGA, a high-speed AD8611 comparator was used to produce the desired output. Unsurprisingly, the TDC measurement results with the SiPM setup were very similar to those attained with the APD.

\end{document}